\documentclass[11pt]{article}
\usepackage[utf8]{inputenc}
\usepackage[T1]{fontenc}
\usepackage{graphicx}
\usepackage{longtable}
\usepackage{wrapfig}
\usepackage{rotating}
\usepackage{amsmath}
\usepackage{amssymb}
\usepackage{capt-of}
\usepackage{hyperref}
\usepackage[margin=1in]{geometry}
\usepackage{mathrsfs}
\usepackage{cancel}
\usepackage{braket}
\usepackage{authblk}

\usepackage{setspace}
\usepackage{xcolor}
\usepackage{amsmath,amsfonts,epsfig}
\usepackage{tcolorbox}
\usepackage{cite}
\usepackage{subcaption}
\usepackage{xstring}
\usepackage[hang,flushmargin]{footmisc} 
\usepackage{enumitem}
\let\svthefootnote\thefootnote
\newcommand\freefootnote[1]{
  \let\thefootnote\relax
  \footnotetext{#1}
  \let\thefootnote\svthefootnote} 

\newcommand\GM[1][]{{\rm GM}^{\IfInteger{#1}{(#1)}{\mathtt{(#1)}}}}
\renewcommand\S[1][]{S^{\IfInteger{#1}{(#1)}{\mathtt{(#1)}}}}

\usepackage{amsmath,amssymb,amsthm,mathtools,amsthm}
\usepackage{bm}
\usepackage{hyperref}

\usepackage{tikz}
\usetikzlibrary{calc}
\usetikzlibrary{arrows.meta}
\usetikzlibrary{shapes.geometric}

\usepackage{graphicx}

\definecolor{colA}{HTML}{378ADD}
\definecolor{colB}{HTML}{EF9F27}
\definecolor{colC}{HTML}{E24B4A}
\definecolor{colD}{HTML}{2C2C2A}
\definecolor{fillket}{HTML}{E1F5EE}
\definecolor{fillbra}{HTML}{FAEAF0}

\newcommand{\Z}{\mathbb Z}

\title{\bf Notes on the Graph-Encoded Manifolds\\ for the (Genuine) Multi-Entropy}
\author[$\spadesuit$,$\heartsuit$]
{Norihiro Iizuka}

\affil[$\spadesuit$]{\it Department of Physics, National Tsing Hua University, Hsinchu 300044, Taiwan}
\affil[$\heartsuit$]{\it Yukawa Institute for Theoretical Physics, Kyoto University, Kyoto 606-8502, Japan}

\date{\today}

\begin{document}

\maketitle

\thispagestyle{empty}
\setcounter{page}{0} 

\vspace{-.3in}
\begin{abstract}
{
Genuine multi-entropy is the part of the $\mathtt{q}$-partite multi-entropy that captures entanglement genuinely shared among all $\mathtt{q}$ parties, rather than entanglement already present among fewer subsystems. It was recently proposed that this genuine part -- so-called a multipartite entanglement signal -- can be studied via the graph-encoded manifold (GEM) built from the underlying $\mathtt{q}$-partite multi-entropy permutation family: reading its $\mathtt{q}$-colored contraction graph $\Gamma_{\mathtt{q},n}$ as a triangulation by $(\mathtt{q}-1)$-simplices. Focusing on $\mathtt{q}=4$, we classify the topology of every vertex link of $\Gamma_{4,n}$ and find $\chi_{\rm link}(n)=n(3-n)$: the graph satisfies the GEM condition, namely every vertex link is a two-sphere, if and only if $n=2$, while for every $n\geq3$ the vertex links are closed surfaces of strictly positive genus $g_n=\tfrac12(n-1)(n-2)$, so that the graph instead defines a simplicial complex with a conical singularity at every vertex.
}

\freefootnote{E-mail: 
\href{mailto:iizuka@phys.nthu.edu.tw}{\texttt{iizuka@phys.nthu.edu.tw}}}
\end{abstract}

\setcounter{footnote}{0}

\newpage


\section{Introduction and Summary}
\label{sec:intro}

Understanding multipartite entanglement beyond correlations that can be
reduced to pairs of subsystems has become an increasingly active problem
in quantum information, quantum many-body physics, and quantum gravity.
A many-body wavefunction contains, in principle, the complete information
about the quantum state, but this information is generally encoded in a
highly complex form. A nontrivial question is how to extract from it
observables that isolate physically meaningful structures of
multipartite entanglement. In particular, one would like to distinguish
entanglement that is genuinely shared among $\mathtt q$ parties from
contributions that can already be attributed to correlations among
fewer subsystems.

A very broad framework for addressing this problem is provided by
\emph{multi-invariants} \cite{Gadde:2023zzj, Gadde:2024taa}. Given several copies of a
multipartite wavefunction and its complex conjugate, one contracts their
indices according to a collection of replica permutations. The resulting
quantities are invariant under independent local unitary transformations
of the subsystems and therefore provide natural probes of multipartite
entanglement. Different choices of replica number and permutation data
define a large class of multi-invariants, from which one may attempt to
extract different aspects of the entanglement structure of the state.

Among these quantities, particularly important examples are those that
treat the participating subsystems symmetrically. For three parties, a
distinguished example is the replica invariant underlying entanglement
negativity, while for general $\mathtt q$, the \emph{multi-entropy}
\cite{Gadde:2022cqi} provides a canonical symmetric construction. The
$\mathtt q$-partite multi-entropy is a natural generalization of the
entanglement R\'enyi entropy and is built from a standard family of
translations acting on $n^{\mathtt q-1}$ replicas of a
$\mathtt q$-partite state. Its symmetric treatment of the subsystems
makes it a natural candidate for probing correlations distributed among
all $\mathtt q$ parties.

The $\mathtt q$-partite multi-entropy itself, however, is not a measure
of genuine $\mathtt q$-partite entanglement. It can respond to
entanglement involving only $\mathtt q-1$ or fewer parties. \emph{Genuine
multi-entropy} was introduced to remove these lower-partite
contributions and isolate the component that is irreducibly associated
with all $\mathtt q$ subsystems \cite{Iizuka:2025ioc,Iizuka:2025caq}.
Explicit constructions are known for $\mathtt q=3,4,5$, and the
properties of these quantities in holography \cite{Iizuka:2025ioc,Iizuka:2025caq,Harper:2024ker,Iizuka:2025elr, Balasubramanian:2025hxg, Balasubramanian:2025jhq, Anegawa:2025prn, Naskar:2026zka, Balasubramanian:2026chr, Fujiki:2026qdt}, QFTs \cite{Harper:2025uui, Berthiere:2025toi}, RTNs and stabilizer systems \cite{Akella:2025owv, Akella:2026bci, Hu:2026bhg, Akella:2026rbe}, 
topological QFTs \cite{Yuan:2025dgx}, and spin systems \cite{Iizuka:2026qqg, Iizuka:2026ahd} are currently under active
investigation. More generally, for multi-invariants that do not treat
all subsystems symmetrically, genuine multipartite contributions can be
organized into so-called genuine multipartite entanglement \emph{signals}
\cite{Gadde:2026msg}. Genuine multi-entropy is the canonical symmetric
example of such a signal.

A particularly interesting geometric interpretation of these signals
was recently proposed in Ref.~\cite{DelZotto:2026fpw}. The authors
conjectured a relation between genuine $\mathtt{q}=(d+1)$-partite
entanglement signals in the ground states of gapped $d$-dimensional
quantum systems and partition functions of the corresponding
long-distance topological quantum field theories on $d$-dimensional
\emph{graph-encoded manifolds (GEMs)}. In this construction, the
colored contraction graph of a multi-invariant is interpreted as a
simplicial complex: each graph vertex represents a $d\,(=\mathtt{q}-1)$-simplex,
and each graph edge of a given color represents the gluing along the
corresponding codimension-one face.\footnote{Concretely, for
$\mathtt{q}=4$ ($d=3$), a graph edge of a given color represents the
triangular face through which two tetrahedra are glued together, and it is not an edge of the tetrahedron itself.} When the appropriate link
conditions are satisfied, the colored graph encodes a closed manifold.
The conjecture then associates the corresponding genuine multipartite
entanglement signal with a TQFT partition function on that manifold.

Applying this correspondence to genuine multi-entropy, the canonical
signal associated with multi-entropy, requires determining the GEM geometry encoded by the multi-entropy contraction graph. This
provides the geometric input needed to apply the proposed signal–TQFT
correspondence to this particularly symmetric family of multi-invariants.
The central goal of this paper is to determine that geometry explicitly. The three-partite case is straightforward, whereas
the four-partite case already exhibits a richer structure. Motivated
also by recent progress in the study of four-partite genuine
multi-entropy \cite{Iizuka:2026ahd,Akella:2026rbe}, we focus primarily
on the four-partite case. Its contraction graph is four-colored and is
interpreted, in the GEM construction, as a simplicial complex assembled
from tetrahedra.

Before stating our results, we stress that the GEM and the boundary
replicated manifold are two distinct geometric constructions associated
with the same colored contraction graph, and that for $\mathtt q=4$ this
difference is one of dimension, not merely of shape. The boundary
replicated manifold is obtained by representing the ket and bra
replicas as $\mathtt q$-gons and gluing them edge-to-edge according to
the replica permutations, producing a two-dimensional replicated
surface for every $\mathtt q$. The GEM construction, by contrast,
interprets each graph vertex as a $(\mathtt q-1)$-simplex and glues
these simplices along their codimension-one faces, producing a
$(\mathtt q-1)$-dimensional simplicial complex whose dimension grows
with $\mathtt q$. For $\mathtt q=3$ the two constructions coincide:
the building block is a triangle in both cases, so the replicated
surface is precisely the simplicial complex itself. For $\mathtt q=4$,
however, the replicated construction still glues quadrilaterals into a
two-dimensional surface, while the GEM construction glues tetrahedra
into a genuinely three-dimensional simplicial complex. Our focus
throughout this paper is entirely on the latter. The boundary
replicated manifold is of separate interest for bulk replica symmetry \cite{Akella:2026bci, Hu:2026bhg} 
and holographic bulk fillings
\cite{Penington:2022dhr,Gadde:2024taa}, and it is reviewed separately in
Appendix~\ref{app:polygon}.

We now turn to the standard permutation family defining the
$\mathtt q=4$ multi-entropy \cite{Gadde:2022cqi}. The replica labels
are elements of the lattice $\mathbb Z_n^3$, where $n$ is the R\'enyi
index\footnote{Throughout this paper, we restrict attention to $n\geq 2$.}, and the four permutations are
\begin{equation}
\pi_A:\bm{x}\mapsto\bm{x}+e_1,\qquad
\pi_B:\bm{x}\mapsto\bm{x}+e_2,\qquad
\pi_C:\bm{x}\mapsto\bm{x}+e_3,\qquad
\pi_D=\mathrm{id},
\label{eq:standardT}
\end{equation}
translating the $i$-th replica coordinate by one unit, $e_i$ the
standard basis vector. As in
Sec.~\ref{sec:gem-construction}, we read this graph as a GEM by
associating a black tetrahedron with each ket replica and a white
tetrahedron with each bra replica, with the color-$X$ face of a white
tetrahedron glued to the color-$X$ face of the black tetrahedron
selected by $\pi_X$. The resulting complex encodes a genuine closed
three-manifold precisely when the link of every vertex -- the closed
surface exposed by cutting out a small neighborhood of that vertex --
is a two-sphere. Determining these vertex links for every $n$, and
hence whether the multi-entropy graph satisfies this GEM
condition, is the goal of this paper.

Our main result is a complete classification of these vertex links for
every R\'enyi index $n$ in the $\mathtt q=4$ multi-entropy family. We
show that each three-colored component obtained by dropping one of the
four colors is isomorphic to the $\mathtt q=3$ replicated manifold with
the same replica index. This determines the topology of every vertex link and, consequently, whether the resulting simplicial complex is a
smooth three-manifold or has conical singularities.

Our main results can be
summarized as follows.
\begin{enumerate}
\item Every vertex link has Euler characteristic
\begin{equation}
\label{eq:chilink}
\chi_{\mathrm{link}}(n)=n(3-n),
\end{equation}
and, since the links are connected and orientable, genus
$g_n=\tfrac12(n-1)(n-2)$.
\item The standard $\mathtt q=4$ graph satisfies the GEM condition if
and only if $n= 2$. For every $n\geq3$, it instead defines a
simplicial complex with $4n$ conical singularities: a torus at $n=3$,
a genus-three surface at $n=4$, and growing genus thereafter.
\end{enumerate}
Thus the standard multi-entropy family that gives a smooth three-sphere at
$n=2$ leaves the manifold sector immediately at $n=3$.

These results sharpen the interpretation of our recent calculation of
four-partite genuine multi-entropy in the toric code
\cite{Akella:2026rbe}. In that work, the $\mathtt q=4$, $n=4$ result was
naturally expected to admit a TQFT interpretation associated with the
corresponding colored graph. The present analysis shows that this graph does not encode a smooth
closed three-manifold, but instead defines a simplicial complex with
conical singularities: its vertex links have genus three. The relevant quantity is therefore expected to be understood,
within the proposed framework of Ref.~\cite{DelZotto:2026fpw}, in terms of a
TQFT amplitude for a manifold with boundary, obtained by removing the
conical singularities, rather than as a conventional closed-manifold
partition function.

The rest of this paper is organized as follows.
Sec.~\ref{sec:GM} reviews the genuine multi-entropy as a signal of multi-entropy. Sec.~\ref{sec:gem-construction} reviews the GEM construction and the
colored-graph dictionary relevant to multi-invariants. Sec.~\ref{sec:q3-lessons} revisits $\mathtt q=3$ cases as a warm-up for $\mathtt q=4$ cases.
Sec.~\ref{sec:gem-analysis} is our main section and  analyzes the cases
$\mathtt q=4$, $n=2,3$, and finally proves the general classification
for arbitrary $n$. 
Sec.~\ref{sec:discussion} discusses the consequences for
multipartite entanglement signals in topological phases. Our convention for the multi-entropy and the review of the replicated
geometry are collected in
Appendices~\ref{app:multientropyrule} and~\ref{app:polygon}.

\section{Genuine multi-entropy}
\label{sec:GM}

To study $\mathtt{q}$-partite entanglement, we will make use of the multi-entropy $S^{(\mathtt q)}$ \cite{Gadde:2022cqi,Penington:2022dhr,Gadde:2023zzj}, which is a natural generalization of
entanglement entropy into $\mathtt{q}$-partite subsystems, and, on top of
it, the genuine multi-entropy $\GM[q]$, which isolates the part of $\S[q]$ that cannot be
reduced to entanglement among fewer than $\mathtt{q}$ parties. For $\mathtt{q}=3$,
\begin{equation}
\begin{aligned}
  \label{GM3}  
  \GM[3](A:B:C) &= \S[3](A:B:C)-\frac{1}{2}\left(S(A)+S(B)+S(C)\right) \\
  &= \S[3](A:B:C) - \left(\mbox{subtraction terms removing bipartite contributions} \right)
\end{aligned}
\end{equation}
For general $\mathtt{q}$, $\GM[q]$ is a fixed linear combination of the
$\mathtt{q}$-partite multi-entropy $\S[q]$ and all lower-partite multi-entropies
$S^{(\tilde{\mathtt{q}})}$, $\tilde{\mathtt{q}}<\mathtt{q}$ \cite{Iizuka:2025ioc, Harper:2024ker}. 

For $\mathtt{q}\geq4$
this combination is not unique, reflecting the existence of several
inequivalent classes of genuine $\mathtt{q}$-partite entanglement \cite{Iizuka:2025caq}. For the
$\mathtt{q}=4$ case studied in this paper, there is one free parameter $a$ \cite{Iizuka:2025ioc, Iizuka:2025caq}:
\begin{equation}
\begin{aligned}
    \label{GM4}
    &\GM[4](A:B:C:D) = \S[4](A:B:C:D) - \frac{1}{3}\left( \S[3](AB:C:D) + \S[3](AC:B:D) + \cdots \right) \\
&\qquad \qquad + a\left(S(AB)+S(AC)+S(AD)\right) + (1/3-a)\left(S(A)+S(B)+S(C)+S(D)\right) \\
&  \,\, = \S[4](A:B:C: D) - \left(\mbox{subtraction terms removing bipartite and tripartite contributions} \right)
\end{aligned}
\end{equation}
where $+\cdots$ stands for summing over all possible permutations of
the boundary subregions (total 6 terms).

For the present paper, only the highest-partite multi-entropy
$\S[\mathtt q]$ is relevant, not the full linear combination defining
$\GM[\mathtt q]$. It is built from the replica invariant
$Z_n^{(\mathtt q)}$ of Eq.~\eqref{eq:replicainvariant}, evaluated on the
standard permutation family, Eq.~\eqref{eq:translations}, acting
on $N=n^{\mathtt q-1}$ replicas. The object of this paper is the
colored contraction graph associated with this standard multi-invariant,
which we interpret as a graph-encoded manifold (GEM) and analyze in the
following sections.


\section{The GEM construction}
\label{sec:gem-construction}

\subsection{Simplices and multi-invariants}

A $k$-simplex is the simplest possible $k$-dimensional figure obtained by
joining $(k+1)$ vertices pairwise.

\begin{center}
\begin{tabular}{c|c|c|c}
Simplex & \# vertices & dimension & example \\
\hline
$0$-simplex & $1$ & $0$ & point \\
$1$-simplex & $2$ & $1$ & edge (line segment) \\
$2$-simplex & $3$ & $2$ & triangle \\
$3$-simplex & $4$ & $3$ & tetrahedron
\end{tabular}
\end{center}

A $\mathtt{q}$-partite state is associated with a $(\mathtt{q}-1)$-simplex (with
$\mathtt{q}$ vertices) \cite{DelZotto:2026fpw}:
\[
\mathtt{q}=3 \ \longrightarrow\ 2\text{-simplex (triangle)}, \qquad
\mathtt{q}=4 \ \longrightarrow\ 3\text{-simplex (tetrahedron)}.
\]
The number $\mathtt{q}$ of parties is thus literally the number of
\emph{faces} of the simplex: each of the $\mathtt{q}$ parties
corresponds to one face, and since every face of a simplex sits
opposite a single vertex, this is equivalent to the more familiar
statement that the simplex has $\mathtt{q}$ vertices. It is this
face-labeling that matters for the construction: the manifold $M$ is
built by gluing many copies of this simplex together, face to face,
matching colors. The picture is the direct $3$-dimensional analogue
of the $\mathtt{q}=3$ case: just as many triangles are glued edge-to-edge
(matching colors) to tile a Riemann surface, many tetrahedra are
glued \emph{face-to-face} (matching colors) to fill out a
$3$-manifold.

For a $\mathtt{q}$-partite state
$|\psi\rangle\in\mathcal H_1\otimes\cdots\otimes\mathcal H_{\mathtt{\mathtt{q}}}$, choosing a
tuple of permutations $(\sigma_1,\dots,\sigma_{\mathtt{\mathtt{q}}})$, $\sigma_k\in S_N$,
defines the multi-invariant
\begin{equation}
Z(\sigma_1,\dots,\sigma_{\mathtt{\mathtt{q}}};|\psi\rangle)
=\langle\psi|^{\otimes N}\,\sigma_1\otimes\cdots\otimes\sigma_{\mathtt{\mathtt{q}}}\,
|\psi\rangle^{\otimes N}.
\end{equation}
This is represented graphically by $N$ black vertices (replicas of
$\psi$) and $N$ white vertices (replicas of $\bar\psi$), with an edge of
color $k$ joining white vertex $a$ to black vertex $\sigma_k(a)$ for each
$k=1,\dots,\mathtt{q}$, producing a bipartite, $\mathtt{q}$-regular, edge-$\mathtt{q}$-colored
graph: it is \emph{bipartite} because every edge runs between a black
and a white vertex only, and it is \emph{$\mathtt{q}$-regular} because every
vertex carries exactly one edge of each of the $\mathtt{q}$ colors. For the
standard multi-entropy family studied here, $N=n^{\mathtt{q}-1}$, with $n$ the R\'enyi
replica index.


We write $\Gamma_{\mathtt{q},n}$ for this graph when $(\sigma_1,\ldots,\sigma_{\mathtt{q}})$ is the standard family at
replica index $n$: a purely combinatorial object, defined for every
$n$, independently of whether it satisfies the GEM condition below.
When it does, $\Gamma_{\mathtt{q},n}$ is the GEM of a genuine closed
manifold $M_{\mathtt{q},n}$; when it does not, no such manifold exists,
and $\Gamma_{\mathtt{q},n}$ instead defines a simplicial complex with
conical singularities.


Putting this together, building a GEM is a simple four-step recipe:
\begin{enumerate}
\item Cut $M$ into pieces that are all simplices (triangles for $\mathtt{q}=3$,
tetrahedra for $\mathtt{q}=4$).
\item Color the pieces themselves black/white, so neighboring pieces are
never the same color.
\item Color the $\mathtt{q}$ faces of \emph{each} piece with $\mathtt{q}$ colors
$1,\dots,\mathtt{q}$, consistently across all pieces.
\item Finally read off a graph, with each piece
becoming a vertex and each gluing along a color-$k$ face becoming an
edge of color $k$.
\end{enumerate}

More precisely, we start with a closed $(\mathtt{q}-1)$-dimensional
manifold $M$, cut into pieces that are all copies of the
$(\mathtt{q}-1)$-simplex. We color the $\mathtt{q}$ faces of each block with
the $\mathtt{q}$ colors $1,\dots,\mathtt{q}$ in a fixed, consistent way, so
that two blocks glued along their color-$k$ face are always glued
face-to-face. We also color the blocks themselves black or white, so
that no two same-colored blocks ever share a glued face -- this is
what we mean by a bipartite triangulation. The resulting bipartite,
$\mathtt{q}$-regular, edge-$\mathtt{q}$-colored graph is what we call the GEM
(graph-encoded manifold) of $(M,\Delta)$.

The previous paragraph went from a manifold $M$ to a graph, by cutting
$M$ into simplices. What we actually need is to run this the other
way: starting from a four-colored graph built out of tetrahedra, we
want to know when it in fact builds a genuine manifold $M$ this way: Given an arbitrary bipartite, $\mathtt{q}$-regular, edge-$\mathtt{q}$-colored
graph, the link of each vertex is built from the subgraph obtained by
dropping one color. We say the graph satisfies the GEM condition if
every such link is a \emph{$(\mathtt{q}-2)$-sphere}. Only then does the graph itself encode a
genuine closed $(\mathtt{q}-1)$-manifold in this way. A graph that fails
this condition instead defines a simplicial complex with conical
singularities, one at each non-spherical vertex link.

\subsection{Case $\mathtt{q}=3$: GEM = boundary replicated manifold}

At $\mathtt{q}=3$, the building block is the $2$-simplex, {\it i.e.}, the
triangle. This is exactly the same building block used in the
replica-polygon construction of Appendix~\ref{app:polygon} (a
$\mathtt{q}$-gon with $\mathtt{q}=3$ is also a triangle). 
So at $\mathtt{q}=3$ there is no distinction to make: the GEM construction and
the boundary replicated manifold $\Sigma_{3,n}$ are the same construction.
Moreover, each two-colored subgraph obtained by dropping one color is
$2$-regular and connected, hence a single cycle, {\it i.e.,} a discrete $S^1$.
Every such graph therefore automatically satisfies the GEM condition.

\subsection{Case \texorpdfstring{$\mathtt{q}=4$}{q=4}: triangulating a 3-manifold with tetrahedra}

Here $M$ is $3$-dimensional, and the building block is the $3$-simplex,
{\it i.e.}, the tetrahedron (4 vertices, 6 edges, 4 triangular faces,
4 colors $A,B,C,D$).

\begin{center}
\begin{tabular}{c|c|c}
Triangulation piece & dim.\ & Graph object \\
\hline
tetrahedron ($3$-simplex) & $3$ & a single graph vertex (black $=\psi$, white $=\bar\psi$) \\
face of the tetrahedron ($2$-simplex) & $2$ & a single edge of one color $\in\{A,B,C,D\}$ \\
edge of the tetrahedron ($1$-simplex) & $1$ & $2$-colored subgraph (a loop) \\
corner of the tetrahedron ($0$-simplex) & $0$ & $3$-colored subgraph (a connected component)
\end{tabular}
\end{center}

\begin{figure}[ht]
\centering
\begin{tikzpicture}[scale=0.85,>=Stealth]
\coordinate (T) at (0,3.6);
\coordinate (L) at (-2.4,1.0);
\coordinate (R) at (2.4,1.0);
\coordinate (F) at (0,-1.4);
\fill[colA!18] (T) -- (L) -- (F) -- cycle;
\fill[colB!22] (T) -- (F) -- (R) -- cycle;
\draw[densely dashed,line width=1pt] (L) -- (R);
\draw[line width=1.3pt] (T) -- (L);
\draw[line width=1.3pt] (T) -- (R);
\draw[line width=1.3pt] (L) -- (F);
\draw[line width=1.3pt] (F) -- (R);
\draw[line width=1.3pt] (T) -- (F);
\coordinate (cL) at ($(T)!0.25!(L)$);
\coordinate (cR) at ($(T)!0.25!(R)$);
\coordinate (cF) at ($(T)!0.25!(F)$);
\draw[colC,densely dashed,line width=1.6pt] (cL) -- (cR);
\draw[colA,densely dashed,line width=1.6pt] (cL) -- (cF);
\draw[colB,densely dashed,line width=1.6pt] (cF) -- (cR);
\node[colA,font=\Large] at (-3.0,1.9) {$A$};
\draw[colA,->] (-2.75,1.75) to[bend right=15] (-1.1,1.0);
\node[colB,font=\Large] at (3.0,1.9) {$B$};
\draw[colB,->] (2.75,1.75) to[bend left=15] (1.0,1.0);
\node[colC,font=\Large] at (1.65,3.55) {$C$};
\draw[colC,->] (1.45,3.4) to[bend left=15] (0.65,3.15);
\node[font=\Large] at (2.9,-1.0) {$D$};
\draw[->] (2.65,-0.85) to[bend left=15] (0.9,-0.15);
\node[font=\small] at (0,-2.1) {tetrahedron};
\begin{scope}[shift={(1.4,5.6)}]
\draw[line width=1.3pt] (-2.4,-1.0) rectangle (3.5,1.0);
\node[font=\normalsize] at (-0.55,0) {vertex link $=$};
\coordinate (t2) at (1.15,0.636);
\coordinate (t3) at (2.75,0.636);
\coordinate (t1) at (1.95,-0.75);
\draw[colC,line width=2.2pt] (t2) -- (t3);
\draw[colA,line width=2.2pt] (t2) -- (t1);
\draw[colB,line width=2.2pt] (t1) -- (t3);
\end{scope}
\draw[->,line width=1.2pt] (0.2,4.55) to[bend right=20] (0,3.75);
\end{tikzpicture}
\caption{
A single tetrahedron in the $\mathtt{q}=4$ GEM. The base $D$ (black)
rests on the ground and is hidden from view; the three side faces
$A$ (blue), $B$ (yellow), and $C$ (red) are visible and meet at the
top vertex. The dashed triangle near the top vertex marks where a
small neighborhood of that vertex is cut off; its three edges lie on
$A$, $B$, $C$ and form the vertex link, shown enlarged (as an
equilateral triangle) in the inset above.
}
\label{fig:tetrahedron-vertexlink}
\end{figure}
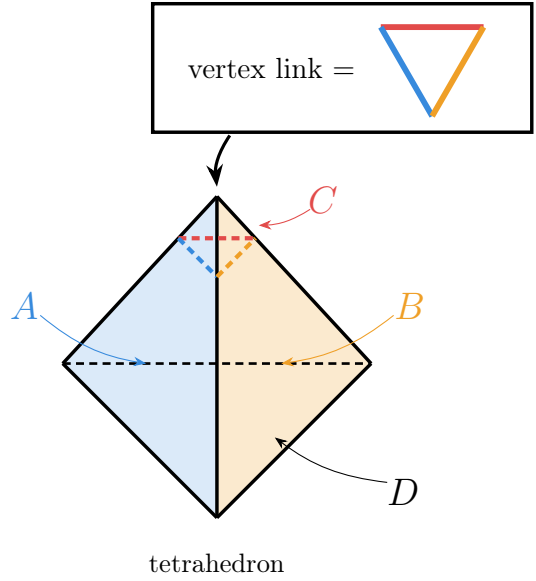

Concretely: one whole tetrahedron $=$ one graph vertex. One of its four
triangular faces $=$ one graph edge of the corresponding color. An edge
of the tetrahedron -- where two faces, say colors $A$ and $C$, meet -- is
tracked by the loops of the $\{A,C\}$ subgraph (keeping those two
colors, dropping $B,D$). A corner (vertex) of the tetrahedron -- where
three faces, say $A,B,C$, meet -- is tracked by the connected components
of the $\{A,B,C\}$ subgraph (dropping just the one remaining color, here
$D$): each such component is itself a $\mathtt{q}=3$-type colored graph, and it
is the link of that corner. The GEM condition at $\mathtt{q}=4$ is precisely
the requirement that every one of these links be a $2$-sphere.
Figure~\ref{fig:tetrahedron-vertexlink} illustrates this for a single
tetrahedron: its base face $D$ rests on the ``ground'' and is hidden,
while the three visible faces $A,B,C$ meet at the top vertex; cutting
off a small neighborhood of that vertex exposes a small triangle whose
three edges lie on the faces $A$, $B$, and $C$ (never $D$, which does
not touch this vertex) -- this triangle is the vertex link.

This observation drives the rest of the
paper.

\section{The \texorpdfstring{$\mathtt{q}=3$}{q=3} case revisited}
\label{sec:q3-lessons}

\subsection{The case \texorpdfstring{$\mathtt{q}=3$, $n=2$}{q=3, n=2}}
\label{sec:q3n2example}

Before turning to $\mathtt{q}=4$, we record the $\mathtt{q}=3$ case explicitly,
since the general formula and the $n=2$ identification below both
reduce to it. Take
\begin{equation}
\sigma_A=(12)(34),\qquad \sigma_B=(13)(24),\qquad \sigma_C=\mathrm{id}.
\end{equation}
Labeling bra replicas $\mathrm{bra}_1,\ldots,\mathrm{bra}_4$ and ket
replicas $\mathrm{ket}_1,\ldots,\mathrm{ket}_4$, and joining
$\mathrm{bra}_a$--$\mathrm{ket}_{\sigma_X(a)}$ for each color $X$, the
full edge correspondence is:
\begin{center}
\begin{tabular}{c|c|c|c}
$a$ & $A$: $\mathrm{bra}_a$--$\mathrm{ket}_{\sigma_A(a)}$
    & $B$: $\mathrm{bra}_a$--$\mathrm{ket}_{\sigma_B(a)}$
    & $C$: $\mathrm{bra}_a$--$\mathrm{ket}_{\sigma_C(a)}$ \\
\hline
$1$ & $\mathrm{bra}_1$--$\mathrm{ket}_2$ & $\mathrm{bra}_1$--$\mathrm{ket}_3$ & $\mathrm{bra}_1$--$\mathrm{ket}_1$ \\
$2$ & $\mathrm{bra}_2$--$\mathrm{ket}_1$ & $\mathrm{bra}_2$--$\mathrm{ket}_4$ & $\mathrm{bra}_2$--$\mathrm{ket}_2$ \\
$3$ & $\mathrm{bra}_3$--$\mathrm{ket}_4$ & $\mathrm{bra}_3$--$\mathrm{ket}_1$ & $\mathrm{bra}_3$--$\mathrm{ket}_3$ \\
$4$ & $\mathrm{bra}_4$--$\mathrm{ket}_3$ & $\mathrm{bra}_4$--$\mathrm{ket}_2$ & $\mathrm{bra}_4$--$\mathrm{ket}_4$ \\
\end{tabular}
\end{center}
This graph is $3$-regular with $F=8$ vertices (four ket and four bra
triangles) and $E=12$ edges (four of each color). Each vertex of this
graph is itself a triangle, and each of its three corners is a vertex
link in the sense of Sec.~\ref{sec:gem-construction}: it is obtained
by dropping one color and tracing the resulting loop of the remaining
two-colored subgraph.

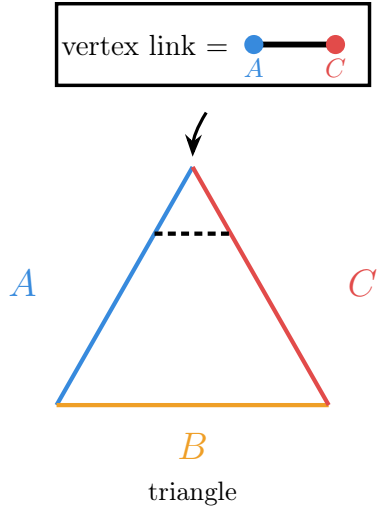
\begin{figure}[t]
\centering
\begin{tikzpicture}[scale=0.9,>=Stealth]
\coordinate (Top) at (0,3.2);
\coordinate (BL)  at (-2.0,-0.3);
\coordinate (BR)  at (2.0,-0.3);

\draw[colA,line width=1.6pt] (Top) -- (BL);
\draw[colB,line width=1.6pt] (BL) -- (BR);
\draw[colC,line width=1.6pt] (Top) -- (BR);

\coordinate (cA) at ($(Top)!0.28!(BL)$);
\coordinate (cC) at ($(Top)!0.28!(BR)$);
\draw[densely dashed,line width=1.6pt] (cA) -- (cC);

\node[colA,font=\Large] at (-2.5,1.5) {$A$};
\node[colB,font=\Large] at (0,-0.9) {$B$};
\node[colC,font=\Large] at (2.5,1.5) {$C$};
\node[font=\small] at (0,-1.6) {triangle};

\begin{scope}[shift={(0,5.0)}]
\draw[line width=1.3pt] (-2.0,-0.6) rectangle (2.6,0.6);
\node[font=\normalsize] at (-0.7,0) {vertex link $=$};
\draw[line width=2.2pt] (0.9,0) -- (2.1,0);
\fill[colA] (0.9,0) circle (4pt);
\fill[colC] (2.1,0) circle (4pt);
\node[colA,font=\small] at (0.9,-0.35) {$A$};
\node[colC,font=\small] at (2.1,-0.35) {$C$};
\end{scope}
\draw[->,line width=1.2pt] (0.2,4.0) to[bend right=15] (0,3.35);
\end{tikzpicture}
\caption{
A single triangle in the $\mathtt{q}=3$ GEM. Cutting off a small
neighborhood of the top vertex (where faces $A$ and $C$ meet) exposes
a short segment with one endpoint on $A$ and one on $C$ -- this is
one edge of the eventual vertex link (a loop, assembled from many
such segments across many triangles). Compare
Fig.~\ref{fig:tetrahedron-vertexlink}, one dimension higher.
}
\label{fig:triangle-vertexlink}
\end{figure}

Dropping color $B$ leaves only the $A$- and $C$-edges. Tracing
alternately along $A$ and $C$ starting from $\mathrm{bra}_1$,
\begin{equation}
\label{eq:vertexlink1}
\mathrm{bra}_1
\xrightarrow{A} \mathrm{ket}_2
\xrightarrow{C} \mathrm{bra}_2
\xrightarrow{A} \mathrm{ket}_1
\xrightarrow{C} \mathrm{bra}_1,
\end{equation}
closes after four steps into the loop
$\{\mathrm{bra}_1,\mathrm{ket}_2,\mathrm{bra}_2,\mathrm{ket}_1\}$; the
remaining four vertices trace out a second loop,
$\{\mathrm{bra}_3,\mathrm{ket}_4,\mathrm{bra}_4,\mathrm{ket}_3\}$.
Repeating this for the other two colors gives, in each case, exactly
two loops of length four:
\begin{equation}
\begin{aligned}
\text{drop }B\ (\{A,C\}):&\quad
\{\mathrm{bra}_1,\mathrm{ket}_2,\mathrm{bra}_2,\mathrm{ket}_1\},
\qquad
\{\mathrm{bra}_3,\mathrm{ket}_4,\mathrm{bra}_4,\mathrm{ket}_3\},
\\
\text{drop }C\ (\{A,B\}):&\quad
\{\mathrm{bra}_1,\mathrm{ket}_2,\mathrm{bra}_4,\mathrm{ket}_3\},
\qquad
\{\mathrm{bra}_2,\mathrm{ket}_1,\mathrm{bra}_3,\mathrm{ket}_4\},
\\
\text{drop }A\ (\{B,C\}):&\quad
\{\mathrm{bra}_1,\mathrm{ket}_3,\mathrm{bra}_3,\mathrm{ket}_1\},
\qquad
\{\mathrm{bra}_2,\mathrm{ket}_4,\mathrm{bra}_4,\mathrm{ket}_2\}.
\end{aligned}
\label{dropBCAq3n2}
\end{equation}
Each loop is a vertex link, and distinct drops give genuinely
different corners of the same triangles, so the counts add:
\begin{equation}
V = \underbrace{2}_{\{A,C\}}+\underbrace{2}_{\{A,B\}}+\underbrace{2}_{\{B,C\}}=6,
\end{equation}
so
\begin{equation}
\chi = V-E+F = 6-12+8=2,\qquad M\cong S^2.
\end{equation}
(See Appendix~\ref{sec:cellcounting}, Eq.~(\ref{eq:q3appendix}) for the
general $\mathtt{q}=3$ formula.)

These six vertices are realized explicitly by the octahedron
(Fig.~\ref{fig:q3n2-octahedron}): the three antipodal vertex pairs are
precisely the $AC$-, $AB$-, and $BC$-type loops computed above, and
its eight triangular faces are the four ket and four bra triangles.

\begin{figure}[t]
\centering
\begin{tikzpicture}[scale=1.0,>=Stealth]
\coordinate (T)  at (0,3);
\coordinate (Bo) at (0,-3);
\coordinate (F)  at (0.6,-1.1);
\coordinate (K)  at (-0.6,1.1);
\coordinate (L)  at (-2.4,0);
\coordinate (R)  at (2.4,0);

\draw[colC,densely dashed,line width=1.3pt] (T) -- (K);
\draw[colC,densely dashed,line width=1.3pt] (Bo) -- (K);
\draw[colB,densely dashed,line width=1.3pt] (R) -- (K);
\draw[colB,densely dashed,line width=1.3pt] (K) -- (L);

\draw[colC,line width=1.6pt] (T) -- (F);
\draw[colC,line width=1.6pt] (Bo) -- (F);
\draw[colA,line width=1.6pt] (T) -- (L);
\draw[colA,line width=1.6pt] (T) -- (R);
\draw[colA,line width=1.6pt] (Bo) -- (L);
\draw[colA,line width=1.6pt] (Bo) -- (R);
\draw[colB,line width=1.6pt] (F) -- (R);
\draw[colB,line width=1.6pt] (L) -- (F);

\coordinate (vL) at ($(T)!0.25!(L)$);
\coordinate (vK) at ($(T)!0.25!(K)$);
\coordinate (vR) at ($(T)!0.25!(R)$);
\coordinate (vF) at ($(T)!0.25!(F)$);

\draw[black,densely dashed,line width=1.6pt]
  (vL) -- (vK) -- (vR) -- (vF) -- cycle;

\foreach \p in {T,Bo,F,K,L,R}{\fill (\p) circle(2pt);}

\node[font=\huge,rotate=30] at (1.2,0.25) {$1$};
\node[font=\huge,rotate=-20] at (-0.6,0.3) {$\bar1$};
\node[font=\huge] at (-0.5,-1.4) {$3$};
\node[font=\huge,rotate=-30] at (0.9,-1.4) {$\bar3$};

\node[font=\large,rotate=-20,text=gray] at (-0.9,1.35) {$2$};
\node[font=\large,text=gray] at (0.65,1.35) {$\bar2$};
\node[font=\large,rotate=-20,text=gray] at (0.65,-0.8) {$4$};
\node[font=\large,rotate=20,text=gray] at (-0.9,-0.8) {$\bar4$};

\node[anchor=north,font=\small,align=left] at (0,-3.5) {
  \textcolor{colA}{\rule{9pt}{9pt}} $A$
  \qquad
  \textcolor{colB}{\rule{9pt}{9pt}} $B$
  \qquad
  \textcolor{colC}{\rule{9pt}{9pt}} $C$
  \quad(dashed = hidden / vertex link)
};
\end{tikzpicture}
\caption{
The standard $\mathtt{q}=3$, $n=2$ GEM is the octahedron: six vertices
(three antipodal pairs), twelve colored edges, and eight triangular
faces. The four visible faces 
are
ket$_1$, bra$_1$, ket$_3$, bra$_3$
($1,\bar1,3,\bar3$), while the four hidden faces through the back
vertex 
are ket$_2$, bra$_2$, ket$_4$, bra$_4$
($2,\bar2,4,\bar4$). The black dashed loop around the top vertex is the vertex link obtained by dropping the $B$-colored edges, corresponding to Eq.~\eqref{eq:vertexlink1}.
}
\label{fig:q3n2-octahedron}
\end{figure}

\subsection{The case \texorpdfstring{$\mathtt{q}=3$, $n=3$}{q=3, n=3}}
\label{sec:q3n3example}

We repeat the same computation for $n=3$, since it will reappear as
the vertex link of the $\mathtt{q}=4$, $n=3$ graph in
Sec.~\ref{sec:n3-detailed}. Take
\begin{equation}
\sigma_A=(123)(456)(789),\qquad
\sigma_B=(147)(258)(369),\qquad
\sigma_C=\mathrm{id}.
\end{equation}
The full edge correspondence, $\mathrm{bra}_a$--$\mathrm{ket}_{\sigma_X(a)}$:
\begin{center}
\begin{tabular}{c|c|c|c}
$a$ & $A$: $\mathrm{bra}_a$--$\mathrm{ket}_{\sigma_A(a)}$
    & $B$: $\mathrm{bra}_a$--$\mathrm{ket}_{\sigma_B(a)}$
    & $C$: $\mathrm{bra}_a$--$\mathrm{ket}_{\sigma_C(a)}$ \\
\hline
$1$ & $\mathrm{bra}_1$--$\mathrm{ket}_2$ & $\mathrm{bra}_1$--$\mathrm{ket}_4$ & $\mathrm{bra}_1$--$\mathrm{ket}_1$ \\
$2$ & $\mathrm{bra}_2$--$\mathrm{ket}_3$ & $\mathrm{bra}_2$--$\mathrm{ket}_5$ & $\mathrm{bra}_2$--$\mathrm{ket}_2$ \\
$3$ & $\mathrm{bra}_3$--$\mathrm{ket}_1$ & $\mathrm{bra}_3$--$\mathrm{ket}_6$ & $\mathrm{bra}_3$--$\mathrm{ket}_3$ \\
$4$ & $\mathrm{bra}_4$--$\mathrm{ket}_5$ & $\mathrm{bra}_4$--$\mathrm{ket}_7$ & $\mathrm{bra}_4$--$\mathrm{ket}_4$ \\
$5$ & $\mathrm{bra}_5$--$\mathrm{ket}_6$ & $\mathrm{bra}_5$--$\mathrm{ket}_8$ & $\mathrm{bra}_5$--$\mathrm{ket}_5$ \\
$6$ & $\mathrm{bra}_6$--$\mathrm{ket}_4$ & $\mathrm{bra}_6$--$\mathrm{ket}_9$ & $\mathrm{bra}_6$--$\mathrm{ket}_6$ \\
$7$ & $\mathrm{bra}_7$--$\mathrm{ket}_8$ & $\mathrm{bra}_7$--$\mathrm{ket}_1$ & $\mathrm{bra}_7$--$\mathrm{ket}_7$ \\
$8$ & $\mathrm{bra}_8$--$\mathrm{ket}_9$ & $\mathrm{bra}_8$--$\mathrm{ket}_2$ & $\mathrm{bra}_8$--$\mathrm{ket}_8$ \\
$9$ & $\mathrm{bra}_9$--$\mathrm{ket}_7$ & $\mathrm{bra}_9$--$\mathrm{ket}_3$ & $\mathrm{bra}_9$--$\mathrm{ket}_9$ \\
\end{tabular}
\end{center}

Dropping color $B$ and tracing from $\mathrm{bra}_1$,
\begin{equation}
\mathrm{bra}_1
\xrightarrow{A}\mathrm{ket}_2
\xrightarrow{C}\mathrm{bra}_2
\xrightarrow{A}\mathrm{ket}_3
\xrightarrow{C}\mathrm{bra}_3
\xrightarrow{A}\mathrm{ket}_1
\xrightarrow{C}\mathrm{bra}_1,
\label{eq:vertexlink2}
\end{equation}
now closes after \emph{six} steps, not four: with $n=3$, a loop must
pass through all $n=3$ distinct index values before returning, so its
length is $2n=6$ rather than $2n=4$. The remaining six vertices give
two more such loops; repeating for the other two colors gives, in
every case, three loops of length six:
\begin{equation}
\begin{aligned}
& \hspace{5mm} \text{drop }B:\, \\
& \{\mathrm{bra}_1,\mathrm{ket}_2,\mathrm{bra}_2,\mathrm{ket}_3,\mathrm{bra}_3,\mathrm{ket}_1\},\ \
\{\mathrm{bra}_4,\mathrm{ket}_5,\mathrm{bra}_5,\mathrm{ket}_6,\mathrm{bra}_6,\mathrm{ket}_4\},\ \
\{\mathrm{bra}_7,\mathrm{ket}_8,\mathrm{bra}_8,\mathrm{ket}_9,\mathrm{bra}_9,\mathrm{ket}_7\},
\\
& \hspace{5mm} \text{drop }C:\,\\
& \{\mathrm{bra}_1,\mathrm{ket}_2,\mathrm{bra}_8,\mathrm{ket}_9,\mathrm{bra}_6,\mathrm{ket}_4\},\ \
\{\mathrm{bra}_2,\mathrm{ket}_3,\mathrm{bra}_9,\mathrm{ket}_7,\mathrm{bra}_4,\mathrm{ket}_5\},\ \
\{\mathrm{bra}_3,\mathrm{ket}_1,\mathrm{bra}_7,\mathrm{ket}_8,\mathrm{bra}_5,\mathrm{ket}_6\},
\\
& \hspace{5mm}\text{drop }A:\, \\
&\{\mathrm{bra}_1,\mathrm{ket}_4,\mathrm{bra}_4,\mathrm{ket}_7,\mathrm{bra}_7,\mathrm{ket}_1\},\ \
\{\mathrm{bra}_2,\mathrm{ket}_5,\mathrm{bra}_5,\mathrm{ket}_8,\mathrm{bra}_8,\mathrm{ket}_2\},\ \
\{\mathrm{bra}_3,\mathrm{ket}_6,\mathrm{bra}_6,\mathrm{ket}_9,\mathrm{bra}_9,\mathrm{ket}_3\}.
\end{aligned}
\end{equation}
So $V=3+3+3=9$, matching $V=\mathtt{q} n^{\mathtt{q}-2}=3\times3$. With
$F=2n^2=18$ and $E=3n^2=27$,
\begin{equation}
\chi=V-E+F=9-27+18=0,\qquad M\cong T^2.
\end{equation}

\begin{figure}[t]
\centering
\resizebox{0.55\textwidth}{!}{%
\begin{tikzpicture}[scale=1.0,>=Stealth]
\def\unit{1.35}
\pgfmathsetmacro{\hgt}{\unit*0.8660254}
\newcommand{\drawcell}[5]{%
  \pgfmathsetmacro{\bx}{0.5*\unit*#1+#2*\unit}
  \pgfmathsetmacro{\bxx}{0.5*\unit*#1+(#2+1)*\unit}
  \pgfmathsetmacro{\by}{-#1*\hgt}
  \pgfmathsetmacro{\tx}{0.5*\unit*(#1+1)+#2*\unit}
  \pgfmathsetmacro{\txx}{0.5*\unit*(#1+1)+(#2+1)*\unit}
  \pgfmathsetmacro{\ty}{-(#1+1)*\hgt}
  \fill[fillket] (\bx,\by) -- (\bxx,\by) -- (\tx,\ty) -- cycle;
  \fill[fillbra] (\bxx,\by) -- (\tx,\ty) -- (\txx,\ty) -- cycle;
  \draw[col#5,line width=1.8pt] (\bxx,\by) -- (\tx,\ty);
  \pgfmathsetmacro{\ketlx}{(\bx+\bxx+\tx)/3}
  \pgfmathsetmacro{\ketly}{(\by+\by+\ty)/3}
  \node[font=\small] at (\ketlx,\ketly) {$#3$};
  \pgfmathsetmacro{\bralx}{(\bxx+\tx+\txx)/3}
  \pgfmathsetmacro{\braly}{(\by+\ty+\ty)/3}
  \node[font=\small] at (\bralx,\braly) {$\bar{#4}$};
}
\drawcell{0}{0}{6}{3}{B}
\drawcell{0}{1}{3}{2}{A}
\drawcell{0}{2}{5}{5}{C}
\drawcell{1}{0}{1}{1}{C}
\drawcell{1}{1}{2}{8}{B}
\drawcell{1}{2}{8}{7}{A}
\drawcell{2}{0}{4}{6}{A}
\drawcell{2}{1}{9}{9}{C}
\drawcell{2}{2}{7}{4}{B}
\foreach \r/\c/\col in {0/0/A,0/1/C,0/2/B,1/0/B,1/1/A,1/2/C,2/0/C,2/1/B,2/2/A}{
  \pgfmathsetmacro{\tx}{0.5*\unit*(\r+1)+\c*\unit}
  \pgfmathsetmacro{\txx}{0.5*\unit*(\r+1)+(\c+1)*\unit}
  \pgfmathsetmacro{\ty}{-(\r+1)*\hgt}
  \draw[col\col,line width=1.8pt] (\tx,\ty) -- (\txx,\ty);
}
\foreach \r/\c/\col in {0/0/C,0/1/B,0/2/A,1/0/A,1/1/C,1/2/B,2/0/B,2/1/A,2/2/C}{
  \pgfmathsetmacro{\bxx}{0.5*\unit*\r+(\c+1)*\unit}
  \pgfmathsetmacro{\by}{-\r*\hgt}
  \pgfmathsetmacro{\txx}{0.5*\unit*(\r+1)+(\c+1)*\unit}
  \pgfmathsetmacro{\ty}{-(\r+1)*\hgt}
  \draw[col\col,line width=1.8pt] (\bxx,\by) -- (\txx,\ty);
}
\foreach \c/\col in {0/C,1/B,2/A}{
  \pgfmathsetmacro{\bx}{\c*\unit}
  \pgfmathsetmacro{\bxx}{(\c+1)*\unit}
  \draw[col\col,line width=1.8pt] (\bx,0) -- (\bxx,0);
}
\foreach \r/\col in {0/A,1/B,2/C}{
  \pgfmathsetmacro{\bx}{0.5*\unit*\r}
  \pgfmathsetmacro{\by}{-\r*\hgt}
  \pgfmathsetmacro{\tx}{0.5*\unit*(\r+1)}
  \pgfmathsetmacro{\ty}{-(\r+1)*\hgt}
  \draw[col\col,line width=1.8pt] (\bx,\by) -- (\tx,\ty);
}

\coordinate (Vlink) at (1.5*\unit,-\hgt);
\def\linkradius{0.30}

\draw[black,densely dashed,line width=1.6pt]
  ($(Vlink)+(0:\linkradius)$)
  -- ($(Vlink)+(60:\linkradius)$)
  -- ($(Vlink)+(120:\linkradius)$)
  -- ($(Vlink)+(180:\linkradius)$)
  -- ($(Vlink)+(240:\linkradius)$)
  -- ($(Vlink)+(300:\linkradius)$)
  -- cycle;

\draw[colA,line width=1.6pt,-{Stealth[length=6pt]}] (0.5*\unit*0.5,-0.5*\hgt) -- ++(-0.45,0);
\draw[colA,line width=1.6pt,-{Stealth[length=6pt]}] (0.5*\unit*0.5+3*\unit,-0.5*\hgt) -- ++(0.45,0);
\draw[colB,line width=1.6pt,-{Stealth[length=6pt]}] (0.5*\unit*1.5,-1.5*\hgt) -- ++(-0.45,0);
\draw[colB,line width=1.6pt,-{Stealth[length=6pt]}] (0.5*\unit*1.5+3*\unit,-1.5*\hgt) -- ++(0.45,0);
\draw[colC,line width=1.6pt,-{Stealth[length=6pt]}] (0.5*\unit*2.5,-2.5*\hgt) -- ++(-0.45,0);
\draw[colC,line width=1.6pt,-{Stealth[length=6pt]}] (0.5*\unit*2.5+3*\unit,-2.5*\hgt) -- ++(0.45,0);
\draw[colC,line width=1.6pt,-{Stealth[length=6pt]}] (0.5*\unit,0) -- ++(0,0.4);
\draw[colC,line width=1.6pt,-{Stealth[length=6pt]}] (0.5*\unit+1.5*\unit,-3*\hgt) -- ++(0,-0.4);
\draw[colB,line width=1.6pt,-{Stealth[length=6pt]}] (\unit+0.5*\unit,0) -- ++(0,0.4);
\draw[colB,line width=1.6pt,-{Stealth[length=6pt]}] (\unit+0.5*\unit+1.5*\unit,-3*\hgt) -- ++(0,-0.4);
\draw[colA,line width=1.6pt,-{Stealth[length=6pt]}] (2*\unit+0.5*\unit,0) -- ++(0,0.4);
\draw[colA,line width=1.6pt,-{Stealth[length=6pt]}] (2*\unit+0.5*\unit+1.5*\unit,-3*\hgt) -- ++(0,-0.4);
\node[anchor=north,font=\small,align=left]
at (2.2*\unit,-3.4*\hgt) {
  \textcolor{colA}{\rule{11pt}{11pt}} $A$
  \qquad
  \textcolor{colB}{\rule{11pt}{11pt}} $B$
  \qquad
  \textcolor{colC}{\rule{11pt}{11pt}} $C$
};
\end{tikzpicture}%
}
\caption{
Explicit triangular-lattice gluing for the standard $\mathtt{q}=3$, $n=3$
multi-entropy permutations, with each internal vertex surrounded by exactly two
alternating colors (an $AC$-, $AB$-, or $BC$-type vertex). The black
dashed hexagonal loop is the vertex link obtained by dropping the
$B$-colored edges, corresponding to Eq.~\eqref{eq:vertexlink2}.
This is the same construction as in Fig.6 of
Ref.\cite{Penington:2022dhr}, up to the convention difference (our convention is $\text{bra}_i\ \longleftrightarrow\ \text{ket}_{i+1}$,
whereas theirs are $\text{bra}_i\ \longleftrightarrow\ \text{ket}_{i-1}$).
The resulting closed surface is $\Gamma_{3,3}\cong T^2$.
}
\label{fig:q3n3-polygon}
\end{figure}
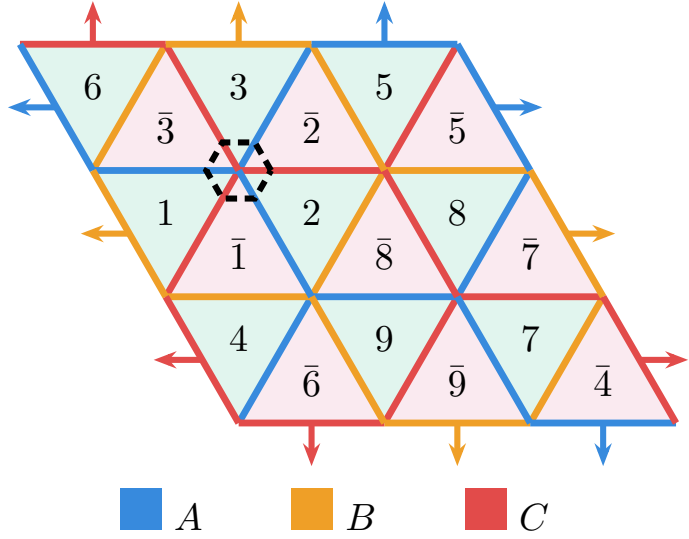

\subsection{Lessons for the four-partite case}
\label{sec:q3-lessons-short}

It is worth stating the logical order explicitly, since it is easy to
conflate its steps -- and the same order will recur, one dimension
higher, throughout Sec.~\ref{sec:generaln}:
\begin{itemize}
\item[1.] The permutations $\sigma_A,\sigma_B,\sigma_C$ alone, purely
algebraically, determine how many vertex links there are, and how
long each loop is (four vertices at $n=2$, six at $n=3$).
\item[2.] Each vertex link is automatically a circle $S^1$: a
two-colored subgraph is $2$-regular, and a connected $2$-regular graph
is always a single cycle -- true regardless of what the ambient
surface $M$ turns out to be, and regardless of $n$. This step alone
never distinguishes $n=2$ from $n=3$.
\item[3.] Only the resulting cell counts $V,E,F$, via $\chi=V-E+F$,
determine the topology of $M$ itself. This is where $n=2$ and $n=3$
genuinely diverge: $\chi=2$ gives $M\cong S^2$ at $n=2$, while
$\chi=0$ gives $M\cong T^2$ at $n=3$.
\item[4.] Only \emph{after} step 3 is it known whether every vertex
link is contractible in $M$: $\pi_1(S^2)$ is trivial, so at $n=2$
every loop is automatically contractible, while $\pi_1(T^2)=\mathbb
Z^2$ is nontrivial, so at $n=3$ the very same kind of loop can fail to
be contractible. This is a consequence of the topology of $M$, not an
independent check, and not a property of the loop in isolation.
\end{itemize}
At $\mathtt{q}=4$, Sec.~\ref{sec:generaln} repeats this logic one
dimension higher -- with one genuinely new step. Step 1 still holds:
permutations determine the vertex links algebraically. Step 2,
however, changes in an essential way: dropping one color now leaves
\emph{three} colors, not two, so each vertex link is a
two-dimensional closed surface rather than a one-dimensional circle.
This is the crucial difference from $\mathtt{q}=3$: a closed
$1$-manifold has no alternative to being $S^1$, but a closed
$2$-manifold does -- $T^2$, genus $2$, and so on -- so sphericity of
the vertex link is no longer automatic. It must be checked, using the
vertex link's \emph{own} cell counts and Euler characteristic (step
2$'$), exactly as $M$ itself was checked in step 3 above. Only once
that check passes is step 3$'$ (spherical vertex links assemble into a
genuine closed $3$-manifold $M$) meaningful, and only then does step
4$'$ (contractibility, now one dimension up) follow. This inserted
check -- whether the vertex link itself is $S^2$ -- is precisely the
GEM condition, and Sec.~\ref{sec:generaln} shows it holds only at
$n=2$; already at $n=3$ the vertex link is $\Gamma_{3,3}\cong T^2$,
not $S^2$.


\section{Topology of \texorpdfstring{$\Gamma_{4,n}$}{Gamma\_{4,n}}}
\label{sec:gem-analysis}

We take the standard multi-entropy permutation family acting on the
$\mathbb Z_n^3$ replica lattice, given by \eqref{eq:standardT}.

\subsection{$\mathtt{q}=4$, $n=2$ case}
\label{sec:q4n2}

For $\mathtt{q}=4$ and $n=2$, the standard replica permutations are
\begin{equation}
\pi_A=(12)(34)(56)(78),\qquad
\pi_B=(13)(24)(57)(68),\qquad
\pi_C=(15)(26)(37)(48),\qquad
\pi_D=\mathrm{id}.
\end{equation}

The replicas are naturally labeled by
\begin{equation}
\bm{x}=(x_1,x_2,x_3)\in\mathbb Z_2^3.
\end{equation}
There are therefore $8$ ket replicas and $8$ bra replicas. We denote them by
\begin{equation}
(\bm{x},0)\equiv\bm{x}_{\rm ket},
\qquad
(\bm{x},1)\equiv\bm{x}_{\rm bra}.
\end{equation}

The four colored edges are defined, from bra to ket, by
\begin{equation}
\begin{aligned}
A:\ &(x_1,x_2,x_3)_{\rm bra}
   \longrightarrow (x_1+1,x_2,x_3)_{\rm ket}, \\
B:\ &(x_1,x_2,x_3)_{\rm bra}
   \longrightarrow (x_1,x_2+1,x_3)_{\rm ket}, \\
C:\ &(x_1,x_2,x_3)_{\rm bra}
   \longrightarrow (x_1,x_2,x_3+1)_{\rm ket}, \\
D:\ &(x_1,x_2,x_3)_{\rm bra}
   \longrightarrow (x_1,x_2,x_3)_{\rm ket},
\label{eq:q4n2-rules}
\end{aligned}
\end{equation}
where all coordinates are understood modulo $2$. When an edge is
traversed from ket to bra, the inverse transformation is used. For
example,
\begin{equation}
(y_1,y_2,y_3)_{\rm ket}
\xrightarrow{\,B\,}
(y_1,y_2+1,y_3)_{\rm bra},
\end{equation}
since subtraction and addition coincide modulo $2$.

To study the vertices of the tetrahedral complex associated with the
color $D$, we drop all $D$-colored edges and retain only the
$\{A,B,C\}$ subgraph. Equivalently, we glue the tetrahedra through
their $A$, $B$, and $C$ faces while leaving their $D$ faces unglued.
The connected components of this three-colored subgraph determine the
vertex links of this type.

Let us define
\begin{equation}
I(\bm{x},\epsilon)
=
x_1+x_2+x_3+\epsilon
\pmod2,
\qquad
\epsilon=
\begin{cases}
0,&\text{ket},\\
1,&\text{bra}.
\end{cases}
\label{eq:I-q4n2}
\end{equation}
Note that this quantity is preserved along every $A$-, $B$-, and $C$-colored
edge, in either direction. For example, along an $A$ edge from bra to
ket,
\begin{equation}
I(x_1+1,x_2,x_3,0)
=
x_1+x_2+x_3+1
=
I(x_1,x_2,x_3,1).
\end{equation}
In the reverse direction,
\begin{equation}
I(y_1+1,y_2,y_3,1)
=
y_1+y_2+y_3
=
I(y_1,y_2,y_3,0),
\end{equation}
where we used $-1\equiv +1\pmod2$. The same calculation applies to the
$B$ and $C$ edges.

This partitions the ket and bra vertices into two sectors, $I=0$ and
$I=1$. Let us display these two sectors explicitly.

For $I=0$, the four ket vertices are
\begin{equation}
(0,0,0),\quad
(0,1,1),\quad
(1,0,1),\quad
(1,1,0),
\label{eq:I0-kets-n2}
\end{equation}
while the four bra vertices are
\begin{equation}
(0,0,1),\quad
(0,1,0),\quad
(1,0,0),\quad
(1,1,1).
\label{eq:I0-bras-n2}
\end{equation}

For $I=1$, the four ket vertices are
\begin{equation}
(0,0,1),\quad
(0,1,0),\quad
(1,0,0),\quad
(1,1,1),
\label{eq:I1-kets-n2}
\end{equation}
while the four bra vertices are
\begin{equation}
(0,0,0),\quad
(0,1,1),\quad
(1,0,1),\quad
(1,1,0).
\label{eq:I1-bras-n2}
\end{equation}
Notice that the ket vertices in the $I=1$ sector are precisely the
bra vertices in the $I=0$ sector, and vice versa. Nevertheless, once
the $D$-colored edges are dropped, the value of $I$ is preserved
along every remaining edge. Hence the $I=0$ and $I=1$ sectors cannot
be connected within the resulting $\{A,B,C\}$ subgraph.

Let us focus on the $I=0$ sector. Inside the $I=0$ ket sector,
\begin{equation}
x_1+x_2+x_3=0\pmod2,
\end{equation}
so that
\begin{equation}
x_3=x_1+x_2.
\end{equation}
Hence every ket vertex is uniquely labeled by
$(x_1,x_2)\in\mathbb Z_2^2$:
\begin{equation}
(x_1,x_2)_{\rm ket}
\longleftrightarrow
(x_1,x_2,x_1+x_2)_{\rm ket}.
\end{equation}

Similarly, for a bra vertex,
\begin{equation}
x_1+x_2+x_3+1=0\pmod2,
\end{equation}
or equivalently
\begin{equation}
x_3=x_1+x_2+1.
\end{equation}
Thus every bra vertex is also uniquely labeled by
$(x_1,x_2)\in\mathbb Z_2^2$.

We can now rewrite the three edge rules entirely in terms of these
two-dimensional labels. The $A$ edge gives
\begin{equation}
(x_1,x_2)_{\rm bra}
\longrightarrow
(x_1+1,x_2)_{\rm ket},
\end{equation}
the $B$ edge gives
\begin{equation}
(x_1,x_2)_{\rm bra}
\longrightarrow
(x_1,x_2+1)_{\rm ket},
\end{equation}
and the $C$ edge gives
\begin{equation}
(x_1,x_2)_{\rm bra}
\longrightarrow
(x_1,x_2)_{\rm ket}.
\end{equation}
Hence
\begin{equation}
\begin{aligned}
A&:\ (x_1,x_2)\longmapsto(x_1+1,x_2),\\
B&:\ (x_1,x_2)\longmapsto(x_1,x_2+1),\\
C&:\ (x_1,x_2)\longmapsto(x_1,x_2).
\end{aligned}
\label{eq:q3n2-rules-inside-q4}
\end{equation}
These are precisely the standard $\mathtt{q}=3$, $n=2$ replica
permutations (bra $\to$ ket), matching the convention of
Sec.~\ref{sec:q3n2example}.

The appearance of a two-dimensional $\mathtt{q}=3$ graph again has a
direct geometric interpretation. The original $\mathtt{q}=4$
contraction graph is regarded as a three-dimensional simplicial
complex assembled from tetrahedra. Removing a small neighborhood of a
vertex replaces each incident tetrahedron by a triangle, whose three
edges inherit the colors $A$, $B$, and $C$. Consequently, the vertex
link is assembled according to exactly the three-colored rules in
Eq.~\eqref{eq:q3n2-rules-inside-q4}.

We have therefore shown, as colored triangulated surfaces, that
\begin{equation}
\text{each $\mathtt{q}=4$, $n=2$ vertex link}
\ \cong\
\Gamma_{3,2},
\label{eq:link-is-gamma32}
\end{equation}
the standard $\mathtt{q}=3$, $n=2$ GEM identified explicitly in
Sec.~\ref{sec:q3n2example} (the octahedron, Fig.~\ref{fig:q3n2-octahedron}).
Since $\chi=2$ there, connected and orientable, this vertex link is $S^2$.

For the choice in which color $D$ is dropped, there are two connected
components, corresponding to $I=0$ and $I=1$. The same argument applies
when any one of the other three colors is dropped. Therefore the
$\mathtt{q}=4$, $n=2$ complex has
\begin{equation}
4\times2=8
\end{equation}
vertex links in total, and every one of them is homeomorphic to
$S^2$. Hence $\Gamma_{4,2}$ satisfies the GEM condition.

As an independent check, the same conclusion follows from cell counts
on the full four-colored graph, without passing through the
$\Gamma_{3,2}$ identification above. There are $F_3=16$ tetrahedra
($n^{\mathtt{q}-1} = 8$ ket $+\,8$ bra); their $16\times4=64$ raw faces glue in pairs to give
$F_2=32$; each of the $\binom{4}{2}=6$ two-colored subgraphs decomposes into
$4$ cycles of length $4$, giving $F_1=24$ \footnote{For colors
$\{A,B\}$ ({\it i.e.},\ dropping $C,D$), this is not a new computation:
restricted to $\{1,2,3,4\}$, $\pi_A,\pi_B$ coincide exactly with
$\sigma_A,\sigma_B$ of Sec.~\ref{sec:q3-lessons}, so the two loops
are precisely the ``drop $C$'' line of
Eq.~\eqref{dropBCAq3n2}. The identical pattern repeats, shifted by
$4$, on $\{5,6,7,8\}$, giving two more loops -- four loops of length
$4$ in total. The same reduction applies to each of the other five
color pairs.}; finally for each of the $4$ ways of dropping one color, the $3$-colored
subgraph decomposes into $2$ components of size $8$, giving
$F_0=4\times2=8$\footnote{Again not a new computation: for $D$, this
is exactly the $I=0$/$I=1$ partition already exhibited above,
Eqs.~\eqref{eq:I0-kets-n2}--\eqref{eq:I1-bras-n2}, where the
invariant $I(\bm{x},\epsilon)$ splits the $16$ vertices into two
sectors of $8$. The same happens for each of the other three colors,
by the identical argument with the roles of $A,B,C,D$ permuted.}, so that
\begin{equation}
\chi=F_0-F_1+F_2-F_3=8-24+32-16=0.
\end{equation}

These cell counts only check a necessary condition for $\Gamma_{4,2}$
to encode a closed $3$-manifold: $\chi=0$ is required but does not by
itself determine which manifold this is. The final consistency check -- whether
each vertex link is actually $S^2$ -- requires examining the internal
structure of a single vertex candidate, which is what we turn to next.

Let us make this concrete, focusing on the $I=0$ component (the same
argument applies verbatim to $I=1$ and to each of the other $7$
vertex candidates).

\paragraph{$F=8$.} The $I=0$ component consists of the four kets
$(0,0,0),(0,1,1),(1,0,1),(1,1,0)$ and the four bras
$(0,0,1),(0,1,0),(1,0,0),(1,1,1)$ listed in
Eqs.~\eqref{eq:I0-kets-n2}--\eqref{eq:I0-bras-n2}: eight triangles in
total. Geometrically (cf.\ Fig.~\ref{fig:tetrahedron-vertexlink}), each of these eight triangles is the $D$-face (or $\bar D$-face, for bra) of one of the eight tetrahedra in the $I=0$ sector -- the face opposite the vertex being cut off, which is exactly the vertex link.

\paragraph{$E=12$.} Each of these eight triangles carries one edge
each of colors $A,B,C$, giving $8\times3=24$ raw edge-ends; each
edge is shared by one bra and one ket, so this halves to $24/2=12$
distinct edges.

\paragraph{$V=6$.} Tracing the loops directly from these coordinates,
dropping $B$ and starting from $\mathrm{bra}(0,0,1)$,
\begin{equation}
\mathrm{bra}(0,0,1)
\xrightarrow{A}\mathrm{ket}(1,0,1)
\xrightarrow{C}\mathrm{bra}(1,0,0)
\xrightarrow{A}\mathrm{ket}(0,0,0)
\xrightarrow{C}\mathrm{bra}(0,0,1),
\label{eq:vertexlink-q4n2}
\end{equation}
closes after four steps; the remaining four vertices,
$\mathrm{bra}(0,1,0),\mathrm{ket}(1,1,0),\mathrm{bra}(1,1,1),
\mathrm{ket}(0,1,1)$, trace out a second loop of length four. This
gives $2$ vertices from dropping $B$. Dropping $A$ or $C$ instead
gives, by the identical computation, $2$ more vertices each, so
\begin{equation}
V=2+2+2=6.
\end{equation}
Hence
\begin{equation}
\chi_{\rm link}=V-E+F=6-12+8=2
\end{equation}
for the $I=0$ component -- and, by the same three computations
applied to each of the other $7$ vertex candidates, for all of them.
Every vertex link is $S^2$.

Geometrically, this is exactly the picture of
Fig.~\ref{fig:tetrahedron-vertexlink}: gluing only through the $A$,
$B$, $C$ faces of several tetrahedra, while leaving their $D$ faces
unglued, produces the small sphere seen when cutting off a
neighborhood of a vertex. Indeed, since $x_3$ is fixed by $x_1,x_2$,
Eq.~\eqref{eq:q3n2-rules-inside-q4} identifies
Eq.~\eqref{eq:vertexlink-q4n2} with Eq.~\eqref{eq:vertexlink1}: the
$V=6$ computation above is not a separate verification but simply
this identification written out in coordinates. Since every one of
the $8$ vertex candidates is $S^2$ in this way, $\Gamma_{4,2}$
satisfies the GEM condition.

\subsection{ \texorpdfstring{$\mathtt{q}=4$, $n=3$}{q=4, n=3} case}
\label{sec:n3-detailed}

We now work out the first singular case, $\mathtt{q}=4$ and $n=3$, in 
detail.  Besides making the appearance of the torus vertex links
explicit, this example exhibits the mechanism behind the general
formula derived in Sec.~\ref{sec:generaln}.

The replicas are labeled by
\begin{equation}
\bm{x}=(x_1,x_2,x_3)\in\mathbb Z_3^3.
\end{equation}
There are therefore $27$ ket replicas and $27$ bra replicas.  We denote
them by
\begin{equation}
(\bm{x},0)\equiv \bm{x}_{\rm ket},
\qquad
(\bm{x},1)\equiv \bm{x}_{\rm bra}.
\end{equation}
The four colored edges are defined, from bra to ket, by
\begin{equation}
\begin{aligned}
A:\ &(x_1,x_2,x_3)_{\rm bra}
   \longrightarrow (x_1+1,x_2,x_3)_{\rm ket}, \\
B:\ &(x_1,x_2,x_3)_{\rm bra}
   \longrightarrow (x_1,x_2+1,x_3)_{\rm ket},  \\
C:\ &(x_1,x_2,x_3)_{\rm bra}
   \longrightarrow (x_1,x_2,x_3+1)_{\rm ket},  \\
D:\ &(x_1,x_2,x_3)_{\rm bra}
   \longrightarrow (x_1,x_2,x_3)_{\rm ket},
\label{eq:q4n3-rules}
\end{aligned}
\end{equation}
where all coordinates are understood modulo $3$.  When an edge is
traversed from ket to bra, the inverse transformation is used.  For
example,
\begin{equation}
(y_1,y_2,y_3)_{\rm ket}
\xrightarrow{\,B\,}
(y_1,y_2-1,y_3)_{\rm bra}.
\end{equation}

To study the vertices of the tetrahedral complex associated with the
color $D$, we drop all $D$-colored edges and retain only the
$\{A,B,C\}$ subgraph.  Equivalently, we glue the tetrahedra through
their $A$, $B$, and $C$ faces while leaving their $D$ faces unglued.
The connected components of this three-colored subgraph determine the
vertex links of this type.


Just as we did in \eqref{eq:I-q4n2} for $n=2$ case in Sec.~ \ref{sec:q4n2}, let us define 
\begin{equation}
I(\bm{x},\epsilon)
=
x_1+x_2+x_3+\epsilon
\pmod 3,
\qquad
\epsilon=
\begin{cases}
0,&\text{ket},\\
1,&\text{bra}.
\end{cases}
\label{eq:I-q4n3}
\end{equation}
The only difference is that everything is now taken modulo 3 instead of modulo 2.
This quantity is preserved along every $A$-, $B$-, and $C$-colored
edge, in either direction.  For example, along an $A$ edge from bra to
ket,
\begin{equation}
I(x_1+1,x_2,x_3,0)
=
x_1+x_2+x_3+1
=
I(x_1,x_2,x_3,1).
\end{equation}
In the reverse direction,
\begin{equation}
I(y_1-1,y_2,y_3,1)
=
y_1+y_2+y_3
=
I(y_1,y_2,y_3,0).
\end{equation}
The same calculation applies to the $B$ and $C$ edges.

Let us display explicitly the sector $I=0$.  The nine ket vertices in
this sector are
\begin{equation}
\begin{aligned}
&(0,0,0),\ (0,1,2),\ (0,2,1),\\
&(1,0,2),\ (1,1,1),\ (1,2,0),\\
&(2,0,1),\ (2,1,0),\ (2,2,2),
\end{aligned}
\label{eq:I0-kets}
\end{equation}
while the nine bra vertices are
\begin{equation}
\begin{aligned}
&(0,0,2),\ (0,1,1),\ (0,2,0),\\
&(1,0,1),\ (1,1,0),\ (1,2,2),\\
&(2,0,0),\ (2,1,2),\ (2,2,1).
\end{aligned}
\label{eq:I0-bras}
\end{equation}

For example, the $A$ edges map the bra list bijectively onto the ket
list:
\begin{equation}
\begin{array}{c@{\quad\xrightarrow{\ A\ }\quad}c}
(0,0,2) &(1,0,2)\\
(0,1,1) &(1,1,1)\\
(0,2,0) &(1,2,0)\\
(1,0,1) &(2,0,1)\\
(1,1,0) &(2,1,0)\\
(1,2,2) &(2,2,2)\\
(2,0,0) &(0,0,0)\\
(2,1,2) &(0,1,2)\\
(2,2,1) &(0,2,1).
\end{array}
\label{eq:A-I0-table}
\end{equation}
The $B$ and $C$ edges likewise map these same nine bra vertices
bijectively onto the same nine ket vertices.  Thus no $A$, $B$, or $C$
edge leaves the $I=0$ sector.  The sectors $I=1$ and $I=2$ behave in
exactly the same way.

This proves that the $\{A,B,C\}$ subgraph splits into three disjoint
sectors, each containing nine ket and nine bra vertices.  It remains to
check that each sector is connected, rather than decomposing further.
This will follow immediately from the explicit identification below
with the standard $\mathtt{q}=3$, $n=3$ graph, which is connected.

\paragraph{Reduction to the \texorpdfstring{$\mathtt{q}=3$, $n=3$}{q=3, n=3}
graph.}

Let us focus on the $I=0$ ket sector. 
Inside the $I=0$ ket sector,
\begin{equation}
x_1+x_2+x_3=0\pmod3,
\end{equation}
so the third coordinate is not independent:
\begin{equation}
x_3=-x_1-x_2\pmod3.
\label{eq:x3-ket-dependent}
\end{equation}
Consequently, every ket vertex in this component is uniquely labeled
by the pair $(x_1,x_2)\in\mathbb Z_3^2$:
\begin{equation}
(x_1,x_2)_{\rm ket}
\ \longleftrightarrow\
(x_1,x_2,-x_1-x_2)_{\rm ket}.
\end{equation}
Similarly, for a bra vertex in the same $I=0$ sector,
\begin{equation}
x_1+x_2+x_3+1=0\pmod3,
\end{equation}
and therefore
\begin{equation}
x_3=-1-x_1-x_2\pmod3.
\label{eq:x3-bra-dependent}
\end{equation}
Thus the bra vertices are also uniquely labeled by
$(x_1,x_2)\in\mathbb Z_3^2$.

Just as previous $\mathtt q=4$, $n=2$ case, we can now rewrite the three edge rules entirely in terms of these
two-dimensional labels.  The $A$ edge gives
\begin{equation}
(x_1,x_2)_{\rm bra}
\longrightarrow
(x_1+1,x_2)_{\rm ket},
\end{equation}
the $B$ edge gives
\begin{equation}
(x_1,x_2)_{\rm bra}
\longrightarrow
(x_1,x_2+1)_{\rm ket},
\end{equation}
and the $C$ edge gives
\begin{equation}
(x_1,x_2)_{\rm bra}
\longrightarrow
(x_1,x_2)_{\rm ket}.
\end{equation}
Hence
\begin{equation}
\begin{aligned}
A&:\ (x_1,x_2)\longmapsto(x_1+1,x_2),\\
B&:\ (x_1,x_2)\longmapsto(x_1,x_2+1),\\
C&:\ (x_1,x_2)\longmapsto(x_1,x_2).
\end{aligned}
\label{eq:q3-rules-inside-q4}
\end{equation}
These are precisely the standard $\mathtt{q}=3$, $n=3$ replica permutations
(bra $\to$ ket, matching the convention of Sec.~\ref{sec:gem-construction}).  Since $x_3$ is uniquely determined by $(x_1,x_2)$ through
Eqs.~\eqref{eq:x3-ket-dependent} and
\eqref{eq:x3-bra-dependent}, it is redundant from this point on. For example, the explicit $x_3$ values in
Eq.~\eqref{eq:A-I0-table} contain no additional information.

The appearance of a two-dimensional $\mathtt{q}=3$ graph here has a direct
geometric meaning.  The original $\mathtt{q}=4$ contraction graph is interpreted
as a three-dimensional complex assembled from tetrahedra.  Cutting off
a small neighborhood of one of its vertices replaces each incident
tetrahedron by a triangle.  The three sides of this triangle inherit
the colors $A$, $B$, and $C$.  The vertex link is therefore a
two-dimensional surface assembled from these triangles according to
exactly the three-colored rules in
Eq.~\eqref{eq:q3-rules-inside-q4}.

We have thus shown, as colored triangulated surfaces, that
\begin{equation}
\text{each $\mathtt{q}=4$, $n=3$ vertex link}
\ \cong\
\Gamma_{3,3},
\label{eq:link-is-gamma33}
\end{equation}
the standard $\mathtt{q}=3$, $n=3$ GEM identified explicitly in
Sec.~\ref{sec:q3n3example} (Fig.~\ref{fig:q3n3-polygon}). Since
$\chi=0$ there, connected and orientable, this vertex link is a
torus: $T^2$.

As an independent check, the same conclusion follows from cell counts
on the full four-colored graph. There are $F_3=54$ tetrahedra
($n^{\mathtt{q}-1}=27$ ket $+\,27$ bra); their $54\times4=216$ raw
faces glue in pairs to give $F_2=108$; each of the $\binom{4}{2}=6$
two-colored subgraphs decomposes into $9$ cycles of length $6$,
giving $F_1=54$; and each of the $4$ ways of dropping one color
leaves a three-colored subgraph with $3$ connected components (the
$I=0,1,2$ sectors above), giving $F_0=12$. Hence
\begin{equation}
\chi=F_0-F_1+F_2-F_3=12-54+108-54=12.
\end{equation}
Unlike at $n=2$, this is \emph{not} zero -- consistent with the fact
that the complex is not actually a closed $3$-manifold at $n=3$: the
formula $\chi=2n(n-1)(n-2)$, obtained by the same counting for
general $n$, vanishes only at $n=2$, exactly the values at which
the GEM condition holds.

An additional check concerns the internal structure of a single vertex
candidate. Repeating the check at $n=3$, focusing on the $I=0$
component (the same argument applies verbatim to $I=1,2$ and to each
of the other $11$ vertex candidates):

\paragraph{$F=18$.} The $I=0$ component consists of the nine kets
and nine bras listed in Eqs.~\eqref{eq:I0-kets}--\eqref{eq:I0-bras}:
eighteen triangles in total, each the $D$-face (or $\bar D$-face)
of one of the eighteen tetrahedra in the $I=0$ sector.

\paragraph{$E=27$.} Each of these eighteen triangles carries one
edge each of colors $A,B,C$, giving $18\times3=54$ raw edge-ends;
each edge is shared by one bra and one ket, so this halves to
$54/2=27$ distinct edges.

\paragraph{$V=9$.} Tracing the loops directly from these coordinates,
dropping $B$ and starting from $\mathrm{bra}(0,0,2)$,
\begin{equation}
\mathrm{bra}(0,0,2)
\xrightarrow{A}\mathrm{ket}(1,0,2)
\xrightarrow{C}\mathrm{bra}(1,0,1)
\xrightarrow{A}\mathrm{ket}(2,0,1)
\xrightarrow{C}\mathrm{bra}(2,0,0)
\xrightarrow{A}\mathrm{ket}(0,0,0)
\xrightarrow{C}\mathrm{bra}(0,0,2),
\end{equation}
closes after six steps; the remaining twelve vertices trace out two
more such loops. This gives $3$ vertices from dropping $B$. Dropping
$A$ or $C$ instead gives, by the identical computation, $3$ more
vertices each, so
\begin{equation}
V=3+3+3=9.
\end{equation}

Hence
\begin{equation}
\chi_{\rm link}=V-E+F=9-27+18=0
\end{equation}
for the $I=0$ component -- and, by the same three computations
applied to each of the other $11$ vertex candidates, for all of
them. Every vertex link is a torus $T^2$, not $S^2$: as at $n=2$,
$x_3$ is fixed by $x_1,x_2$, so Eq.~\eqref{eq:q3-rules-inside-q4}
identifies the loop above with the first ``drop $B$'' loop of
Sec.~\ref{sec:q3n3example}, $\mathrm{bra}_1\to\mathrm{ket}_2\to
\mathrm{bra}_2\to\mathrm{ket}_3\to\mathrm{bra}_3\to\mathrm{ket}_1\to
\mathrm{bra}_1$; the $V=9$ computation above is simply this
identification written out in coordinates. Since every one of the
$12$ vertex candidates fails to be $S^2$ in this way, the
$\mathtt{q}=4$, $n=3$ complex fails the GEM condition, with twelve
conical singularities, each locally modeled on the cone $C(T^2)$.


\subsection{General formula for all \texorpdfstring{$n$}{n}}
\label{sec:generaln}

We now derive, rather than merely observe case by case, the exact
link topology for arbitrary $n$.

\paragraph{Step 1: dropping color $D$.} Replicas are indexed by
$\bm x=(x_1,x_2,x_3)\in\mathbb Z_n^3$; colors $A,B,C$ act as translations
by $e_1,e_2,e_3$. Every one of these three colors sends a white replica
$\bm x$ to a black replica $\bm x+e_i$, and in every case the coordinate
sum increases by exactly $1$:
\begin{equation}
s(\bm x+e_i)\equiv s(\bm x)+1\pmod n,\qquad s(\bm x):=x_1+x_2+x_3\bmod n.
\end{equation}
More precisely, writing each vertex as $(\bm x,\epsilon)\in\mathbb Z_n^3
\times\mathbb Z_2$ with $\epsilon=0$ black, $\epsilon=1$ white, the
quantity
\begin{equation}
I(\bm x,\epsilon):=x_1+x_2+x_3+\epsilon\pmod n
\end{equation}
is exactly invariant along every $A$-, $B$-, or $C$-colored edge --
this is precisely the invariant of Eqs.~\eqref{eq:I-q4n2} and
\eqref{eq:I-q4n3} above, now for general $n$. The (keep $\{A,B,C\}$)
subgraph therefore splits into exactly $n$ components, $s=0,\ldots,n-1$,
each consisting of the $n^2$ black replicas with $s(\bm x)=s$ together
with the $n^2$ white replicas with $s(\bm y)=s-1$, for a component
size $F=2n^2$.

\paragraph{Step 2: each component is the standard $\mathtt{q}=3$ family.}
Parametrize black replicas of one component by $(x_1,x_2)\in\mathbb
Z_n^2$ (with $x_3=s-x_1-x_2$ determined) and white replicas by their own
$(y_1,y_2)$. Tracking where each color sends $(y_1,y_2)$:
\begin{equation}
A:\ (y_1,y_2)\mapsto(y_1+1,y_2),\quad
B:\ (y_1,y_2)\mapsto(y_1,y_2+1),\quad
C:\ (y_1,y_2)\mapsto(y_1,y_2).
\end{equation}
This is \emph{exactly} the standard $\mathtt{q}=3$ multi-entropy
family -- the same reduction carried out explicitly for $n=2,3$ in
Secs.~\ref{sec:q4n2} and~\ref{sec:n3-detailed} above, now for
arbitrary $n$. Applying the cell-counting of Appendix~\ref{app:polygon}
(Eq.~\eqref{eq:FEV}) with $\mathtt{q}=3$ and this same replica
index $n$, each component has $F=2n^2,\ E=3n^2,\ V=3n$, hence
\begin{equation}
\chi_{\rm link}(n)=V-E+F=3n-3n^2+2n^2=n(3-n).
\end{equation}

\paragraph{Step 3: dropping color $A$, $B$, or $C$.} The same
conclusion holds for the other three ways of dropping a color, by an
identical argument with the roles of the colors permuted: keeping
$\{B,C,D\}$ (say), the component label is now $x_1\bmod n$ rather than
$x_1+x_2+x_3+\epsilon$ -- since $D$ is the identity, a $D$-edge connects
$(\bm x,0)$ directly to $(\bm x,1)$ at fixed $\bm x$, while $B$- and
$C$-edges shift $x_2,x_3$ freely -- but this again partitions the graph
into exactly $n$ components of size $2n^2$, each isomorphic as a colored
graph to the standard $\mathtt{q}=3$ family (now with $(B,C,D)$ playing the
roles of the coordinate shifts $(e_1,e_2,\mathrm{id})$). The same
cell-counting gives the same 
$F=2n^2,\ E=3n^2,\ V=3n$, hence the same
$\chi_{\rm link}(n)=n(3-n)$. Dropping $B$ or $C$ instead of $A$ follows
by the color relabeling $A\leftrightarrow B$ or $A\leftrightarrow C$.

Thus in conclusion, every one of the $4n$ vertex-link components,
for any of the four ways of dropping a color, is isomorphic as a
colored graph to the standard $\mathtt{q}=3$ family, giving the uniform result
\begin{equation}
\chi_{\rm link}(n)=n(3-n)
\end{equation}
for every one of the $4n$ vertex-link components. Since a closed
orientable surface is a two-sphere precisely when its Euler
characteristic is two, the standard $\mathtt{q}=4$ graph satisfies the
GEM condition only for $n=2$. For every $n\geq3$, each vertex link
instead has genus
\begin{equation}
g_n=\frac{(n-1)(n-2)}{2},
\end{equation}
so that the standard $\mathtt{q}=4$ graph defines a closed three-manifold
only for $n=2$.

\subsubsection{The singularities worsen quickly as we increase $n$} The two cases already
worked out in detail, $n=2$ ($\chi_{\rm link}=2$, $S^2$, Sec.~\ref{sec:q4n2})
and $n=3$ ($\chi_{\rm link}=0$, $T^2$, Sec.~\ref{sec:n3-detailed}),
are the two mildest points on this curve. One replica step further,
at $n=4$, the formula already gives
\begin{equation}
\chi_{\rm link}(4)=4(3-4)=-4,\qquad g_4=\frac{3\times2}{2}=3:
\end{equation}
every one of the $16$ vertex-link components is a genus-$3$ surface
-- strictly worse than the torus at $n=3$, and, unlike $n=2,3$,
already too complicated to draw explicitly by hand. Table~\ref{tab:gem-summary}
summarizes all three cases.

\begin{table}[t]
\centering
\begin{tabular}{c|c|c|c}
$n$ & link topology & $\chi_{\rm link}$ & GEM condition? \\
\hline
$2$ & $S^2$ & $2$ & satisfied \\
$3$ & $T^2$ & $0$ & fails ($12$ conical points) \\
$4$ & genus-$3$ surface & $-4$ & fails ($16$ conical points)
\end{tabular}
\caption{The standard $\mathtt{q}=4$ vertex-link topology for
$n=2,3,4$, from $\chi_{\rm link}(n)=n(3-n)$. The singularities strictly
worsen with each step past $n=2$.}
\label{tab:gem-summary}
\end{table}

$\vspace{-10mm}$
\section{Discussion}
\label{sec:discussion}

As shown above, the standard $\mathtt{q}=4$ multi-entropy graph defines
a smooth three-manifold only for $n=2$. Starting at $n=3$, the
associated simplicial complex instead develops conical singularities.
According to the proposed dictionary of
Ref.~\cite{DelZotto:2026fpw}, in particular their Eq.~(4.20), one should
excise a small neighborhood of each such singularity before assigning
a TQFT interpretation. The genuine four-partite signal for $n\geq3$
should therefore no longer be interpreted as
$Z_{\rm TQFT}(M)$ for a closed three-manifold. Rather, following their
proposal, it is expected to involve TQFT amplitudes on the resulting
manifold with boundary, with the canonical Dirichlet boundary
condition imposed on each boundary component. Our classification determines the boundary topology entering this
proposed interpretation; it does not by itself establish the
corresponding TQFT relation for genuine multi-entropy as a signal.

In the present construction, excising the $4n$ singular vertices
produces $4n$ boundary components, each homeomorphic to
$\Sigma_{g_n}$, where
$g_n=\frac12(n-1)(n-2)$.
Thus, for example, the $n=3$ geometry has twelve torus boundary
components, whereas the $n=4$ geometry has sixteen boundary
components of genus three.

This observation also provides a natural geometric interpretation of
the $\mathtt{q}=4,n=4$ toric-code result of
Ref.~\cite{Akella:2026rbe}. The new topological information identified
there cannot be interpreted simply as $Z_{\rm TQFT}(M)$ for a closed
three-manifold, since the underlying standard graph no longer
satisfies the GEM manifold condition: each vertex link has genus $3$.
Instead, following the general framework of Ref.~\cite{DelZotto:2026fpw},
it is associated with a TQFT quantity defined on a manifold with
Dirichlet boundaries.

It is also noteworthy that the onset of the singular regime and the
onset of genuinely new four-partite information do not coincide. The
standard graph first ceases to represent a smooth manifold at
$n=3$, where the genuine multi-entropy still collapses to the
tripartite information $I_3$. Only at $n=4$, one replica step
later, does this collapse fail. 
This separation suggests that, at least in this example, the
appearance of conical singularities is a geometric precursor to, but
not by itself sufficient for, genuinely new topological information
to emerge.

More broadly, this classification suggests a shift in perspective:
for the standard $\mathtt{q}=4$ permutation family, a smooth closed manifold occurs only at $n=2$,
whereas the singular, Dirichlet-boundary case is generic for $n\geq3$
and thus occurs for infinitely many replica indices. Whether the
genuine multi-entropy as a signal of multi-entropy indeed computes the corresponding
Dirichlet-boundary TQFT quantity predicted by
Ref.~\cite{DelZotto:2026fpw} remains an interesting open question.
Testing this proposal directly within the replica construction is an
interesting direction for future work.

$\vspace{-10mm}$
\section*{Acknowledgements}
We thank Sriram Akella and Akihiro Miyata for collaboration on related work
\cite{Akella:2026rbe} and for helpful comments on
the manuscript.
The work of N.I. was supported in part by MEXT KAKENHI Grant-in-Aid for
Transformative Research Areas A ``Extreme Universe'' No.\ 21H05184. The
work of N.I. was also supported in part by NSTC of Taiwan Grant Number
114-2112-M-007-025-MY3.

\appendix 

\section{Multi-entropy rule}
\label{app:multientropyrule}

The direction convention used throughout this paper -- an edge of
color $k$ joins the bra replica $\bm x$ to the ket replica
$g_k(\bm x)$, rather than the other way around -- is not a free choice
but is forced by the definition of the replica trick itself. For a
bipartite density matrix $\rho_{ab}=\psi_a\psi^*_b$ built from a pure
state, the $n$-th Rényi replica invariant is
\begin{equation}
\mathrm{Tr}(\rho^n)
=\sum \rho_{a_1a_2}\rho_{a_2a_3}\cdots\rho_{a_na_1}
=\sum(\psi_{a_1}\psi^*_{a_2})(\psi_{a_2}\psi^*_{a_3})\cdots(\psi_{a_n}\psi^*_{a_1}).
\label{eq:trace-rule}
\end{equation}
Copy $i$ in this product contributes the factor
$\psi_{a_i}\psi^*_{a_{i+1}}$: its ket carries index $a_i$ and its bra
carries index $a_{i+1}$. The trace contracts $a_{i+1}$ with the ket
index of copy $i+1$, so it is the \emph{bra} of copy $i$ that is
directly identified with the \emph{ket} of copy $i+1$,
\begin{equation}
\text{bra}_i\ \longleftrightarrow\ \text{ket}_{i+1},
\label{eq:bra-ket-rule}
\end{equation}
and not the reverse. Figure~\ref{fig:q3n3-multientropy} below, and the
edge-direction conventions of Secs.~\ref{sec:gem-construction}
and~\ref{sec:n3-detailed}, follow this rule.

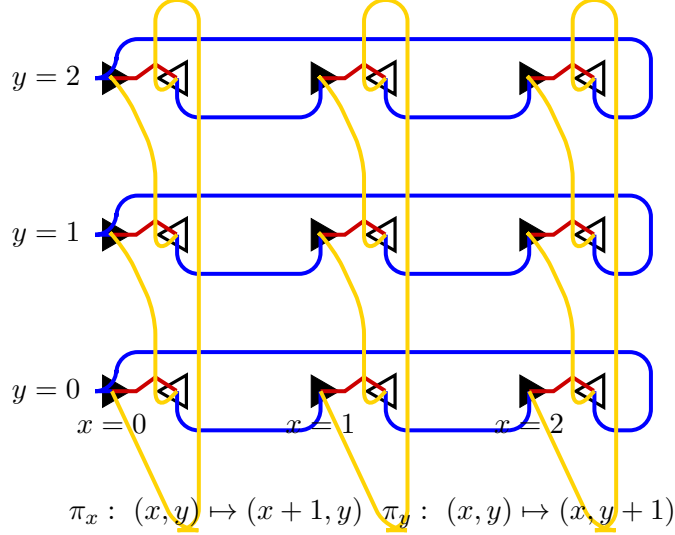
\begin{figure}[t]
\centering
\begin{tikzpicture}[
    scale=1.15,
    ket/.style={
        regular polygon,
        regular polygon sides=3,
        shape border rotate=270,
        fill=black,
        draw=black,
        minimum size=14pt,
        inner sep=0pt
    },
    bra/.style={
        regular polygon,
        regular polygon sides=3,
        shape border rotate=90,
        fill=white,
        draw=black,
        line width=1.2pt,
        minimum size=14pt,
        inner sep=0pt
    },
    xedge/.style={
        blue,
        line width=1.5pt
    },
    yedge/.style={
        yellow!70!orange,
        line width=1.5pt
    },
    local/.style={
        red!80!black,
        line width=1.5pt
    }
]

\def\dx{2.4}
\def\dy{1.8}

\foreach \i in {0,1,2}{
    \foreach \j in {0,1,2}{

        \coordinate (K\i\j) at (\i*\dx,\j*\dy);
        \coordinate (B\i\j) at (\i*\dx+0.75,\j*\dy);

        \node[ket] at (K\i\j) {};
        \node[bra] at (B\i\j) {};

        \draw[local]
        (K\i\j)
        -- ++(0.28,0)
        -- ++(0.22,0.16)
        -- ++(0.25,-0.16)
        -- (B\i\j);
    }
}

\foreach \j in {0,1,2}{
    \draw[xedge,rounded corners=8pt]
    (B0\j)
    -- ++(0,-0.45)
    -- ($(K1\j)+(0,-0.45)$)
    -- (K1\j);

    \draw[xedge,rounded corners=8pt]
    (B1\j)
    -- ++(0,-0.45)
    -- ($(K2\j)+(0,-0.45)$)
    -- (K2\j);

    \draw[xedge,rounded corners=8pt]
    (B2\j)
    -- ++(0,-0.45)
    -- ++(0.65,0)
    -- ++(0,0.9)
    -- ++(-6.15,0)
    -- ++(0,-0.45)
    -- (K0\j);
}

\foreach \i in {0,1,2}{
    \draw[yedge,rounded corners=8pt]
    (B\i0)
    -- ++(-0.25,-0.25)
    -- ++(0,1.15)
    -- ($(K\i1)+(0.15,-0.15)$)
    -- (K\i1);

    \draw[yedge,rounded corners=8pt]
    (B\i1)
    -- ++(-0.25,-0.25)
    -- ++(0,1.15)
    -- ($(K\i2)+(0.15,-0.15)$)
    -- (K\i2);

    \draw[yedge,rounded corners=8pt]
    (B\i2)
    -- ++(-0.25,-0.25)
    -- ++(0,1.15)
    -- ++(0.5,0)
    -- ++(0,-6.1)
    -- ++(-0.25,0)
    -- (K\i0);
}

\node[below=10pt] at (1.2,-0.8)
{$\pi_x:\ (x,y)\mapsto(x+1,y)$};

\node[below=10pt] at (4.8,-0.8)
{$\pi_y:\ (x,y)\mapsto(x,y+1)$};

\node[left=8pt] at (K02) {$y=2$};
\node[left=8pt] at (K01) {$y=1$};
\node[left=8pt] at (K00) {$y=0$};

\node[below=5pt] at (K00) {$x=0$};
\node[below=5pt] at (K10) {$x=1$};
\node[below=5pt] at (K20) {$x=2$};

\end{tikzpicture}
\caption{
Schematic contraction pattern for the standard
$\mathtt{q}=3$, $n=3$ multi-entropy $S^{(\mathtt q=3)}_{n=3}$.
Black and open vertices denote ket and bra replicas, respectively.
The blue and yellow lines represent the two cyclic replica
permutations $\pi_x$ and $\pi_y$.
}
\label{fig:q3n3-multientropy}
\end{figure}

\section{The polygonal replica construction}
\label{app:polygon}

This appendix reviews the two-dimensional replica-polygon construction
of \cite{Penington:2022dhr},
applied to the standard multi-entropy permutation family, together with its Euler
characteristic formula and the underlying cell-counting arguments.
\subsection{The GEM manifold and the replicated manifold}
\label{sec:gem-vs-replica}

The manifold $M$ of the GEM construction (Sec.~\ref{sec:gem-construction})
and the replicated manifold of this appendix should not be confused,
even though both are built from the same colored contraction graph.
The GEM construction starts from $M$ itself, triangulates it into
$d$-simplices, and reads off the graph as a byproduct: manifold first,
graph second. The replicated manifold runs the other way: it starts
from the replica permutations, glues ket and bra copies accordingly,
and only then does a manifold emerge. For $\mathtt{q}=3$ these two
directions happen to agree, since a $2$-simplex is already a triangle.
For $\mathtt{q}=4$ they do not: the same graph vertex is read as a
tetrahedron on the GEM side and as a quadrilateral on the replica
side, so there is no reason for the two resulting geometries to
coincide -- and indeed, as we saw in the main text, they do not. The
boundary replicated manifold reviewed here is of separate interest
for bulk replica symmetry \cite{Akella:2026bci,Hu:2026bhg} and
holographic bulk fillings \cite{Penington:2022dhr,Gadde:2024taa}; we
do not pursue that direction further here.

\subsection{Replica invariant and cyclic junction}

Consider a pure state $\ket{\psi}\in \mathcal H_{A_1}\otimes\cdots\otimes\mathcal H_{A_{\mathtt{\mathtt{q}}}}$.
The R\'enyi-$n$, $\mathtt{q}$-partite multi-entropy is built from the replica
invariant
\begin{equation}
Z_n^{(\mathtt{q})}=
\bra{\psi}^{\otimes n^{\mathtt{q}-1}}
P_1(g_1)P_2(g_2)\cdots P_{\mathtt{\mathtt{q}}}(g_{\mathtt{\mathtt{q}}})
\ket{\psi}^{\otimes n^{\mathtt{q}-1}},
\label{eq:replicainvariant}
\end{equation}
where the replicas are labeled by $\bm{x}=(x_1,\ldots,x_{\mathtt{q}-1})\in\Z_n^{\mathtt{q}-1}$,
i.e.\ $N=n^{\mathtt{q}-1}$ in the notation of Sec.~\ref{sec:gem-construction}.
For the standard multi-entropy family,
\begin{align}
g_i\cdot\bm{x}&=\bm{x}+\bm e_i \pmod n,
&& i=1,\ldots,\mathtt{q}-1,
\label{eq:translations}\\
g_{\mathtt{\mathtt{q}}}&=e.
\end{align}
One permutation can be gauge-fixed to the identity; we choose $g_{\mathtt{\mathtt{q}}}=e$.

We arrange the $\mathtt{q}$ regions cyclically, $A_1-A_2-\cdots-A_{\mathtt{\mathtt{q}}}-A_1$.
Following \cite{Penington:2022dhr}, each ket copy and each bra copy is represented by a
$\mathtt{q}$-gon; there are $n^{\mathtt{q}-1}$ ket and $n^{\mathtt{q}-1}$ bra polygons, and the edge
of type $i$ belonging to bra replica $\bm x$ is glued to the edge of
type $i$ belonging to ket replica $g_i(\bm x)$. The resulting
two-dimensional cell complex is the link surrounding the
minimal multipartite junction; we call it the \emph{boundary replicated
manifold} $\Sigma_{\mathtt{q},n}$. Filling the junction amounts to taking
the cone $C(\Sigma_{\mathtt{q},n})$, which is locally manifold-like at its apex
iff $\Sigma_{\mathtt{q},n}\cong S^2$.

\subsection{Cell counting and the Euler characteristic}
\label{sec:cellcounting}

For the standard $\mathtt{q}$-partite construction, the number of replicas is
\begin{equation}
n^{\mathtt{q}-1}.
\end{equation}
Since there is one ket face and one bra face for each replica, the total
number of $\mathtt{q}$-gonal faces is
\begin{equation}
F=2n^{\mathtt{q}-1}.
\end{equation}
Each face has $\mathtt{q}$ edges, so before gluing there are
$\mathtt{q}\times2n^{\mathtt{q}-1}$ edge segments.
Since every edge is identified with exactly one partner, the number of
edges in the quotient is
\begin{equation}
E=\frac{\mathtt{q}\times2n^{\mathtt{q}-1}}{2}
=
\mathtt{q} n^{\mathtt{q}-1}.
\end{equation}
Similarly, each $\mathtt{q}$-gon has $\mathtt{q}$ corners, giving
$\mathtt{q}\times2n^{\mathtt{q}-1}$ corners before identification.
For each of the $\mathtt{q}$ vertex types, the replica gluing identifies
$2n$ corners into a single quotient vertex.
Hence the total number of vertices is
\begin{equation}
V=
\frac{\mathtt{q}\times2n^{\mathtt{q}-1}}{2n}
=
\mathtt{q} n^{\mathtt{q}-2}.
\label{eq:FEV}
\end{equation}
Therefore the Euler characteristic of the boundary replicated manifold
$\Sigma_{\mathtt{q},n}$ is
\begin{align}
\chi(\Sigma_{\mathtt{q},n})
&=
F-E+V
\nonumber\\
&=
2n^{\mathtt{q}-1}
-\mathtt{q} n^{\mathtt{q}-1}
+\mathtt{q} n^{\mathtt{q}-2}
\nonumber\\
&=
n^{\mathtt{q}-2}\bigl[\mathtt{q}-(\mathtt{q}-2)n\bigr].
\label{eq:chi}
\end{align}
For $\mathtt{q}=3$, this reduces to
\begin{equation}
\chi(\Sigma_{3,n})=n(3-n),
\label{eq:q3appendix}
\end{equation}
agreeing with the tripartite result of \cite{Penington:2022dhr}

The same general direction was already pursued in
Ref.~\cite{Gadde:2024taa}, where a multi-invariant defined by replica
permutations is associated with a closed replicated manifold, and its
topology is analyzed through the cycle structure of the relative
permutations. Their Eq.~(3.8), a Riemann--Hurwitz-type formula,
\begin{equation}
\chi(M_E)
=
n_r
\left[
\chi(M)
-
\sum_{\sigma}
\left(
1-\frac{1}{k_\sigma}
\right)
\right],
\label{eq:GHK-RH}
\end{equation}
holds when all cycles of a twist permutation $\sigma$ have the same
length $k_\sigma$, with $M$ the original unreplicated manifold, $M_E$
the replicated manifold, and $n_r$ the number of replicas. For the
standard $\mathtt{q}$-partite family, $M=S^2$, $\chi(M)=2$,
$n_r=n^{\mathtt{q}-1}$, $k_\sigma=n$, with $\mathtt{q}$ relative twist
permutations; substituting into Eq.~\eqref{eq:GHK-RH} reproduces
Eq.~\eqref{eq:chi} exactly. The cell-counting above is thus a direct
specialization of their Riemann--Hurwitz analysis to the standard multi-entropy 
family.

\bibliographystyle{JHEP}
\bibliography{refs}

@article{Akella:2026rbe,
    author = "Akella, Sriram and Iizuka, Norihiro and Miyata, Akihiro",
    title = "{Genuine Multi-Entropy in the Toric Code}",
    eprint = "2607.06050",
    archivePrefix = "arXiv",
    primaryClass = "hep-th",
    month = "7",
    year = "2026"
}

@article{Fujiki:2026qdt,
    author = "Fujiki, Kosei and Tasuki, Kenya",
    title = "{Multi-entropy in heavy local quenches}",
    eprint = "2606.12526",
    archivePrefix = "arXiv",
    primaryClass = "hep-th",
    reportNumber = "YITP-26-65",
    month = "6",
    year = "2026"
}

@article{Hu:2026bhg,
    author = "Hu, Miao and Lin, Simon and Nechita, Ion",
    title = "{Multi-entropy in random tensor networks}",
    eprint = "2606.04470",
    archivePrefix = "arXiv",
    primaryClass = "hep-th",
    month = "6",
    year = "2026"
}

@article{Balasubramanian:2026chr,
    author = "Balasubramanian, Vijay and Chan, William K. L. and Kang, Monica Jinwoo and Murdia, Chitraang and Ross, Simon F.",
    title = "{Constraints on four-party entanglement in holography}",
    eprint = "2606.00210",
    archivePrefix = "arXiv",
    primaryClass = "hep-th",
    month = "5",
    year = "2026"
}

@article{Naskar:2026zka,
    author = "Naskar, Joydeep",
    title = "{On a mixed-state extension of the holographic signal inequality}",
    eprint = "2605.26617",
    archivePrefix = "arXiv",
    primaryClass = "hep-th",
    month = "5",
    year = "2026"
}

@article{Akella:2026bci,
    author = "Akella, Sriram and Iizuka, Norihiro",
    title = "{Structural Obstruction to Replica Symmetry Breaking for Multi-Entropy in Random Tensor Networks}",
    eprint = "2604.13261",
    archivePrefix = "arXiv",
    primaryClass = "hep-th",
    month = "4",
    year = "2026"
}

@article{Iizuka:2026ahd,
    author = "Iizuka, Norihiro and Miyata, Akihiro",
    title = "{The Junction Law for Multipartite Entanglement in Confining Holographic Backgrounds}",
    eprint = "2604.10583",
    archivePrefix = "arXiv",
    primaryClass = "hep-th",
    month = "4",
    year = "2026"
}

@article{Gadde:2026msg,
    author = "Gadde, Abhijit",
    title = "{On genuine multipartite entanglement signals}",
    eprint = "2603.07680",
    archivePrefix = "arXiv",
    primaryClass = "quant-ph",
    month = "3",
    year = "2026"
}

@article{Iizuka:2026qqg,
    author = "Iizuka, Norihiro and Miyata, Akihiro",
    title = "{Where Multipartite Entanglement Localizes: The Junction Law for Genuine Multi-Entropy}",
    eprint = "2602.16331",
    archivePrefix = "arXiv",
    primaryClass = "hep-th",
    month = "2",
    year = "2026"
}

@article{DelZotto:2026fpw,
    author = "Del Zotto, Michele and Gadde, Abhijit and Putrov, Pavel",
    title = "{From Multipartite Entanglement to TQFT}",
    eprint = "2602.16770",
    archivePrefix = "arXiv",
    primaryClass = "hep-th",
    month = "2",
    year = "2026"
}

@article{Anegawa:2025prn,
    author = "Anegawa, Takanori and Suzuki, Shota and Tamaoka, Kotaro",
    title = "{Black Holes as a Multipartite Entanglers: Multientropy in AdS3/CFT2}",
    eprint = "2512.21037",
    archivePrefix = "arXiv",
    primaryClass = "hep-th",
    doi = "10.1093/ptep/ptag047",
    journal = "PTEP",
    volume = "2026",
    number = "4",
    pages = "043B03",
    year = "2026"
}

@article{Balasubramanian:2025jhq,
    author = "Balasubramanian, Vijay and Jiang, Hanzhi and Ross, Simon F.",
    title = "{Time evolution of multi-party entanglement signals}",
    eprint = "2511.16729",
    archivePrefix = "arXiv",
    primaryClass = "hep-th",
    doi = "10.1007/JHEP06(2026)055",
    journal = "JHEP",
    volume = "06",
    pages = "055",
    year = "2026"
}

@article{Yuan:2025dgx,
    author = "Yuan, Ma-Ke and Li, Mingyi and Zhou, Yang",
    title = "{Multi-entropy from Linking in Chern-Simons Theory}",
    eprint = "2510.18408",
    archivePrefix = "arXiv",
    primaryClass = "hep-th",
    month = "10",
    year = "2025"
}

@article{Akella:2025owv,
    author = "Akella, Sriram",
    title = "{Tripartite entanglement in the HaPPY code is not holographic}",
    eprint = "2510.08520",
    archivePrefix = "arXiv",
    primaryClass = "hep-th",
    month = "10",
    year = "2025"
}

@article{Balasubramanian:2025hxg,
    author = "Balasubramanian, Vijay and Kang, Monica Jinwoo and Cummings, Charlie and Murdia, Chitraang and Ross, Simon F.",
    title = "{Purely Greenberger-Horne-Zeilinger{\textendash}like Entanglement is Forbidden in Holography}",
    eprint = "2509.03621",
    archivePrefix = "arXiv",
    primaryClass = "hep-th",
    doi = "10.1103/g5rw-nvnr",
    journal = "Phys. Rev. Lett.",
    volume = "136",
    number = "3",
    pages = "031602",
    year = "2026"
}

@article{Berthiere:2025toi,
    author = "Berthi{\`e}re, Cl{\'e}ment and Gaudin, Paul",
    title = "{Genuine multientropy, dihedral invariants, and Lifshitz theory}",
    eprint = "2509.00593",
    archivePrefix = "arXiv",
    primaryClass = "hep-th",
    doi = "10.1103/vcqd-rkmn",
    journal = "Phys. Rev. D",
    volume = "113",
    number = "6",
    pages = "065029",
    year = "2026"
}

@article{Iizuka:2025elr,
    author = "Iizuka, Norihiro and Miyata, Akihiro and Nishida, Mitsuhiro",
    title = "{Multipartite Markov gaps and entanglement wedge multiway cuts}",
    eprint = "2507.15262",
    archivePrefix = "arXiv",
    primaryClass = "hep-th",
    doi = "10.1007/JHEP10(2025)148",
    journal = "JHEP",
    volume = "10",
    pages = "148",
    year = "2025"
}

@article{Harper:2025uui,
    author = "Harper, Jonathan and Mollabashi, Ali and Takayanagi, Tadashi and Tasuki, Kenya",
    title = "{Multientropy and the dihedral measures at quantum critical points}",
    eprint = "2506.10396",
    archivePrefix = "arXiv",
    primaryClass = "hep-th",
    reportNumber = "YITP-25-87",
    doi = "10.1103/67k4-vc74",
    journal = "Phys. Rev. Res.",
    volume = "7",
    number = "4",
    pages = "043194",
    year = "2025"
}

@article{Iizuka:2025caq,
    author = "Iizuka, Norihiro and Lin, Simon and Nishida, Mitsuhiro",
    title = "{More on genuine multientropy and holography}",
    eprint = "2504.16589",
    archivePrefix = "arXiv",
    primaryClass = "hep-th",
    doi = "10.1103/x76v-mr6n",
    journal = "Phys. Rev. D",
    volume = "112",
    number = "6",
    pages = "066014",
    year = "2025"
}

@article{Harper:2024ker,
    author = "Harper, Jonathan and Takayanagi, Tadashi and Tsuda, Takashi",
    title = "{Multi-entropy at low Renyi index in 2d CFTs}",
    eprint = "2401.04236",
    archivePrefix = "arXiv",
    primaryClass = "hep-th",
    reportNumber = "YITP-24-02",
    doi = "10.21468/SciPostPhys.16.5.125",
    journal = "SciPost Phys.",
    volume = "16",
    number = "5",
    pages = "125",
    year = "2024"
}

@article{Penington:2022dhr,
    author = "Penington, Geoff and Walter, Michael and Witteveen, Freek",
    title = "{Fun with replicas: tripartitions in tensor networks and gravity}",
    eprint = "2211.16045",
    archivePrefix = "arXiv",
    primaryClass = "hep-th",
    doi = "10.1007/JHEP05(2023)008",
    journal = "JHEP",
    volume = "05",
    pages = "008",
    year = "2023"
}

@article{Gadde:2023zzj,
    author = "Gadde, Abhijit and Krishna, Vineeth and Sharma, Trakshu",
    title = "{Towards a classification of holographic multi-partite entanglement measures}",
    eprint = "2304.06082",
    archivePrefix = "arXiv",
    primaryClass = "hep-th",
    doi = "10.1007/JHEP08(2023)202",
    journal = "JHEP",
    volume = "08",
    pages = "202",
    year = "2023"
}

@article{Iizuka:2025ioc,
    author = "Iizuka, Norihiro and Nishida, Mitsuhiro",
    title = "{Genuine multientropy and holography}",
    eprint = "2502.07995",
    archivePrefix = "arXiv",
    primaryClass = "hep-th",
    doi = "10.1103/714c-byxq",
    journal = "Phys. Rev. D",
    volume = "112",
    number = "2",
    pages = "026011",
    year = "2025"
}

@article{Gadde:2022cqi,
    author = "Gadde, Abhijit and Krishna, Vineeth and Sharma, Trakshu",
    title = "{New multipartite entanglement measure and its holographic dual}",
    eprint = "2206.09723",
    archivePrefix = "arXiv",
    primaryClass = "hep-th",
    reportNumber = "TIFR/TH/22-34",
    doi = "10.1103/PhysRevD.106.126001",
    journal = "Phys. Rev. D",
    volume = "106",
    number = "12",
    pages = "126001",
    year = "2022"
}

@article{Gadde:2024taa,
    author = "Gadde, Abhijit and Harper, Jonathan and Krishna, Vineeth",
    title = "{Multi-invariants and bulk replica symmetry}",
    eprint = "2411.00935",
    archivePrefix = "arXiv",
    primaryClass = "hep-th",
    reportNumber = "YITP-24-130, LCTP-24-19",
    doi = "10.1007/JHEP06(2025)116",
    journal = "JHEP",
    volume = "06",
    pages = "116",
    year = "2025"
}

\end{document}